\documentclass[letterpaper,twocolumn,10pt]{article}
\usepackage{url}
\usepackage[compact]{titlesec}
\usepackage{usenix,epsfig,endnotes}
\usepackage{subfiles}
\usepackage[dvipsnames,table]{xcolor}
\usepackage{amssymb,amsthm,amsmath}
\usepackage{soul}
\usepackage{bm}
\usepackage[font=bf]{caption}
\usepackage{multirow}

\usepackage{pdfpages}

\usepackage{xspace}

\newcommand{\Webservicelong}{Web  service\xspace}

\newcommand{\Webservice}{WS\xspace}
\newcommand{\Webservices}{WSes\xspace}

\newcommand{\WS}{WS\xspace}
\newcommand{\WSES}{WSes\xspace}

\newcommand{\Thirdpartyservice}{TPS\xspace}
\newcommand{\Thirdpartyservices}{TPSes\xspace}

\newcommand{\TPS}{TPS\xspace}

\newcommand{\DoS}{DoS\xspace}
\newcommand{\RoQ}{RoQ\xspace}
\newcommand{\RoS}{RoS\xspace}

\newcommand{\name}[1]{\mathtt{#1}}
\newcommand{\orderedlist}[1]{\langle #1 \rangle}

\newcounter{insightlabel}
\newcounter{insightnmbr}
\renewcommand{\theinsightlabel}{\textbf{\theinsightnmbr}}
\newenvironment{insight}{
\begin{list}{\textbf{Observation }\theinsightlabel:~}{\usecounter{insightlabel}\stepcounter{insightnmbr}\setlength{\labelwidth}{0pt}\setlength{\labelsep}{0pt}\setlength{\leftmargin}{0in}\noindent\rule{\linewidth}{1pt}\vspace{-6pt}\item \bf \em}}{\\[-6pt]\end{list}\vspace{-6pt}\noindent\rule{\linewidth}{1pt}}

\newcommand{\mypara}[1]{\medskip\noindent{\bf {#1}:}~}

\newcounter{packednmbr}

\newenvironment{packeditemize}{
\begin{list}{$\bullet$}{
    \setlength{\itemsep}{0pt}
    \addtolength{\labelwidth}{10pt}
    \setlength{\leftmargin}{12pt}
    \setlength{\listparindent}{\parindent}
    \setlength{\parsep}{2pt}
    \setlength{\topsep}{0pt}}}
    {\end{list}
}

\begin{document}

%don't want date printed
\date{}

%make title bold and 14 pt font (Latex default is non-bold, 16 pt)
\title{\vspace{-2cm}\Large \bf Oh, What a Fragile Web We Weave:  Third-party Service Dependencies In Modern Webservices and Implications \\ 
% \normalsize{PaperId: 475}\vspace{-2cm}
}

\author{
{\rm Aqsa Kashaf}\\
Carnegie Mellon University
\and
{\rm Carolina Zarate}\\
Carnegie Mellon University
\and
{\rm Hanruo Wang}\\
Carnegie Mellon University
\and
{\rm Yuvraj Agarwal}\\
Carnegie Mellon Universtiy
\and
{\rm Vyas Sekar}\\
Carnegie Mellon Universtiy
}

\maketitle

% Use the following at camera-ready time to suppress page numbers.
% Comment it out when you first submit the paper for review.
\thispagestyle{empty}

\begin{abstract}
The recent October 2016 DDoS attack on Dyn served as a wakeup call to the security community as many popular and independent  webservices (e.g., Twitter, Spotify) were impacted. This incident raises a larger question on the fragility of  modern webservices due to their dependence on  third-party services. In this paper, we characterize the  dependencies of popular webservices on third party services and how these can lead to \DoS, \RoQ attacks and reduction in  security posture. In particular, we focus on three critical infrastructure services: DNS, CDNs, and certificate authorities (CAs). We analyze both direct relationships (e.g., Twitter uses Dyn) and  indirect dependencies (e.g.,Netflix uses Symantec as OCSP and Symantec in turn uses Verisign for DNS). 

Our key findings are: (1) 73.14\% of the top 100,000 popular services are vulnerable to reduction in availability due to potential attacks on third-party DNS, CDN, CA services that they exclusively rely on; (2) the use of third-party services is concentrated, so that if the top-10 providers of CDN, DNS and OCSP services go down, they can potentially impact 25\%-46\% of the top 100K most popular web services;  
(3) transitive dependencies significantly increase the set of webservices that exclusively depend on popular CDN and DNS service providers, in some cases by ten times (4) targeting even less popular webservices can potentially cause significant collateral damage, affecting upto 20\% of the top-100K webservices due to their shared dependencies. 
  Based on our findings, we present a number of key implications and guidelines to guard against such Internet-scale incidents in the future.
 
\end{abstract}

\section{Introduction}

In October 2016, a major managed DNS provider, Dyn, fell victim to a Distributed Denial of Service (DDoS) attack~\cite{dyn}. Dyn hosted the authoritative nameservers for many popular web services, such as PayPal, Twitter, and Github and as a result they were inaccessible to a sizable part of the East Coast. It turned out that many other webservices were also dependent on Dyn, therefore the incident resulted in a massive impact on the availability of many popular webservices. 

This recent incident suggests that the modern webservices ecosystem is perhaps much more fragile than one would expect, and that threat actors are already taking advantage of this.  At the same time, this incident also raises many broader questions about the robustness (and fragility) of the web ecosystem:
 
 \begin{packeditemize}
  \item How robust or fragile are popular webservices in terms of their dependency on third-party infrastructure services? E.g., do they use any form of redundancy when using third-party services?
 
 \item Is the Dyn incident a singular type of occurrence or there are other types of hidden third-party services that are also potential Achilles' heel for affecting popular webservices? 
 
 \item Are there more subtle or transitive dependencies between webservices and their third-party providers; e.g.,  loading  webservice Netflix entails third-party service Symantec which in turn depends on Verisign for DNS

 \end{packeditemize}

In this paper, we systematically address these questions using a measurement-driven approach by analyzing the top 100,000 sites (as defined by the Alexa rankings)~\cite{alexa}. We believe that such a study is timely and significant to understand the security of modern webservices. While there are some key limitations of the types of end-to-end vantage point measurement studies (e.g., we do not have capacity estimates for third-party services, we cannot infer third-party dependencies that are not visible to end hosts, we don't consider inter-webservice dependencies), we believe that our findings are nevertheless valuable and relevant. The answers to the aforementioned questions have key implications both for different stakeholders as we shed light on key bottlenecks that are potential vectors for future large-scale attacks.

Specifically, we focus on three key types of third-party services that most modern webservices rely on and that are critical pieces in the lifetime of a web request: naming (DNS), SSL certificate validation (OCSP), and content delivery (CDN). 
We analyze two kinds of potential dependencies: (1)  {\em direct} dependencies of the simple kind found in the Dyn incident where a webservice like Spotify uses Dyn as its DNS provider and (2) {\em indirect} or transitive dependencies that consider ``multi-hop'' effects; e.g., loading Netflix.com entails loading {Symantec} which in turn depends on Verisign for DNS.

Our key findings are: 
	\begin{packeditemize}
    \item 73.14\% of the top 100,000 popular services are vulnerable to reduction in availability due to potential attacks on third-party DNS, CDN, CA services that they exclusively rely on;
    \item The use of third-party services is concentrated, so that if the top-10 providers of CDN, DNS and OCSP services go down, they can potentially impact 25\%-46\% of the most popular web services;
    \item Transitive dependencies significantly increase the set of webservices that exclusively depend on popular CDN and DNS service providers, in some cases by ten times, and even affect the set of most popular providers;
    \item Transitive dependencies can introduce even simpler attack capabilities to achieve similar goals. For example, targeting less popular webservices can potentially cause significant collateral damage, affecting upto 20\% of the top-100K webservices due to their shared dependencies. 

    \end{packeditemize}

Based on these findings, we derive key implications 
for future services and attacks. Specifically, we recommend that: (1)  webservices seek to increase their robustness by adding more redundancy w.r.t third party services and also 
be aware of the hidden dependencies of the third-party services they employ and (2) Third-party services increase the transparency in reporting potential attack events and also provide a quantitative understanding of their infrastructure and dependencies to the web services. Finally, we also observe that these have implications for attackers as they can uncover new attack targets for indirectly affecting webservices that can maximize the impact for a fixed amount of attack resources.

\section{Background and Motivation}

\label{sec:background}
In this section we begin with real-world case studies that highlight three potential threats that webservices (\Webservices) may face as a result of the dependencies with other third-party services (\Thirdpartyservices). We specifically focus on three types of potential threats in this paper as discussed below.

\paragraph{Denial of Service (DoS)}
Denial of Service (DoS) occurs when users of a particular service are unable to access it, for reasons such as an attack on it or an internal failure. In 2016, a Distributed Denial of Service (DDoS) attack on a Managed DNS provider called Dyn caused the unavailability of many websites such as Twitter, Spotify, Github etc.\cite{dyn} Dyn is the authoritative name server of these and many other websites. Authoritative name servers are responsible for name resolutions to a particular zone where a DNS zone is a contiguous namespace for which administrative responsibility has been delegated to a single authority. For example, if a user goes to twitter.com then the authoritative name server for twitter will be responsible for answering queries to twitter.com which in this scenario was Dyn. If the authoritative name server becomes inaccessible as a result of an attack or a bug, it can lead to failed name resolutions for all the domains that rely on it, which makes the clients unable to access those domains. This is what happened in this case where a dependency of multiple \Webservices on a single provider led to denial of service for all of their users. 

Another such incident happened with Cloudflare, a major DNS and CDN provider, when a software bug caused DNS and HTTP requests of some of its customer websites to fail \cite{cloudflareLeap}. Similarly, in 2013, Cloudflare effectively disappeared from the Internet due to a router misconfiguration. These incidents show that the dependency of websites on various services become single points of failure in the face of an attack or misconfiguration \cite{cloudflare2013}. 

\paragraph{Reduction of Security (RoS)}
Reduction of Security happens when the security of a \Webservice is compromised, as a result of an attack or misconfiguration. To give an example of such an attack, consider the public-private key cryptography primitives used along with Certificate Authorities (CA's) used to provide end-to-end confidentiality (encryption) and authentication on the Internet. CA's signed certificates presented by web services, which clients verify to be valid. CAs can sometimes revoke certificates when they are compromised and one way to do that is to provide an OCSP service, which clients can query to check the validity of a certificate presented by a website. In a recent ROS attack in 2016, users were unable to connect securely to websites which used GlobalSign as their CA, since their OCSP servers mistakenly marked valid certificates for numerous websites as invalid due to a misconfiguration \cite{globalsignOCSP}.

\paragraph{Reduction of Quality (RoQ)}
Reduction of Quality happens when a \Webservice falls victim to performance degradation, resulting in increased page load times for its users as a consequence of different types of throttling or admission control mechanisms~\cite{roq}. This is not good for a number of reasons, 1) Users eventually stop using the service when they experience poor performance, 2) This results in loss of revenue for the \Webservices.  There can be many ways to do a Reduction of Quality attack. For instance,  Stark et~al.,\cite{stark2012case} describes the additional delay that occurs in the TLS handshake when a user validates the status of a certificate by contacting the OCSP server. They show that OCSP validation can cause a mean delay of approximately 500ms. So if these OCSP requests go into timeout, due to high load or unavailability of the OCSP server, it can cause huge performance penalty for the users.
Moreover, a certain form of denial of service attacks can also cause huge performance degradations \cite{roq} which involves sending a sustained traffic workload that keeps the \Webservice flooded with traffic, without causing DoS but performance degradation to its users.
Such kind of attacks, can also exploit the various \Thirdpartyservice dependencies that exist in the Internet infrastructure.

\paragraph{Motivating Questions}
These case studies and incidents spark several natural questions about the state of the web services ecosystem today:

\begin{packeditemize}

\item How pervasive are such dependencies between popular \Webservices and various \Thirdpartyservices (e.g., DNS, CDN) providers? 

\item How many of these dependencies are direct and how many of them are hidden or indirect transitive dependencies? 

\item How robust or fragile is a typical \Webservice today with respect to different types of \Thirdpartyservices?

\item Are there other service providers (such as Dyn) that are key points of vulnerability on the Internet, such that an attack on them could lead to significant collateral damage to multiple webservices?

\end{packeditemize}

Our goal is to shed light on these questions to better understand the robustness/fragility of the web ecosystem and to provide recommendations for improving their robustness going forward. In the next section, we describe our measurement methodology to collect data about the service dependencies on the Alexa's \cite{alexa} top 100,000 \Webservices and then present our analysis and implications.

% \subfile{background}

\section{Preliminaries and Methodology}

\label{sec:methodology}
In this section, we describe our measurement methodology to shed light on the motivating questions listed above. We also define the types of dependencies we focus on in our analysis. We also highlight key limitations of our analysis to help put our results in context.

\subsection{Problem Scope and Definitions}

\begin{figure}[!h]
\centering
\includegraphics[width=1\columnwidth]{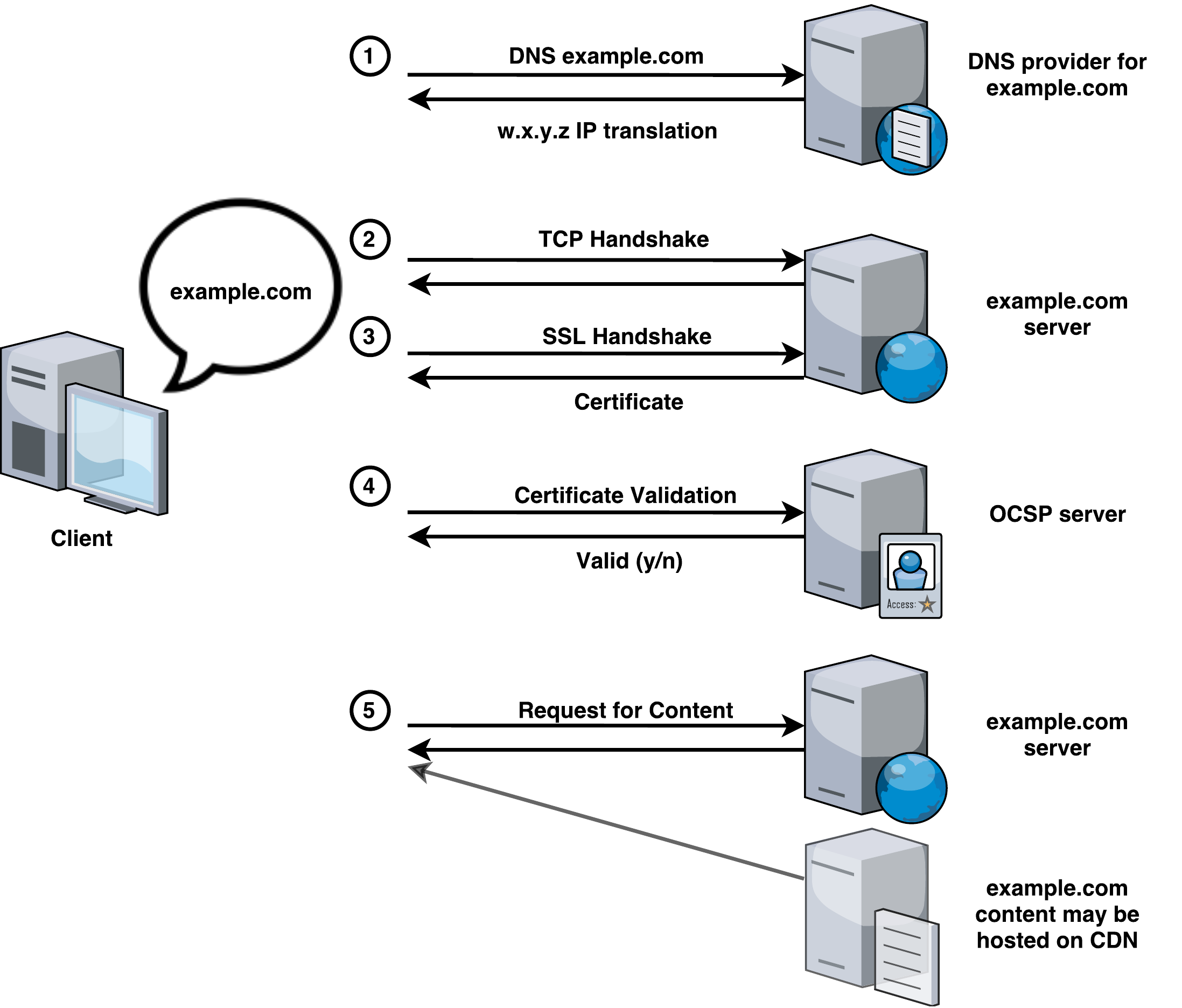}
\caption{The figure shows the complete life cycle of a web request and the different services it interacts with.} 
\label{fig:webreq}
\end{figure}

 To understand the types of services we need to analyze, it is  instructive to start by looking at the life cycle of a typical web request as illustrated in Figure \ref{fig:webreq}. When a user makes request to a \Webservicelong (\Webservice) such as example.com, the request first goes through a name resolution phase, where the hostname to IP translation is fetched from the authoritative nameserver of example.com. After having the successful resolution, the request then goes through a routing phase to reach example.com server. If the website is using HTTPS, the request goes through a certificate exchange phase. The client then consults the respective OCSP server to validate the certificate. After this, user requests the homepage of example.com and the content of example.com might be hosted by a CDN, and loading the page may entail contacting one or more \Thirdpartyservices  as well.

 We refer to the origin service (e.g., example.com) as a \Webservices and non-origin hosted services as \Thirdpartyservices (which can be either infrastructure services such as CDN, DNS, or CA or other \Webservices).

We can model the observed relationships between the various  services as a dependency graph:
$G=(S,D)$, where \emph{S} is the set of services and \emph{D} is the set of directed edges depicting a dependency between services. Each node in the graph is represented as a tuple: $s=\orderedlist{\name{name}, \name{service \; type}}$, 
where \emph{service type} can be a website, CDN, DNS or OCSP. Edges  represent s dependencies, denoted as: $d = \orderedlist{\name{source}, \name{target}, \name{dependency \; type}}$. 

\begin{figure}[!h]
\centering
\includegraphics[width=1\columnwidth]{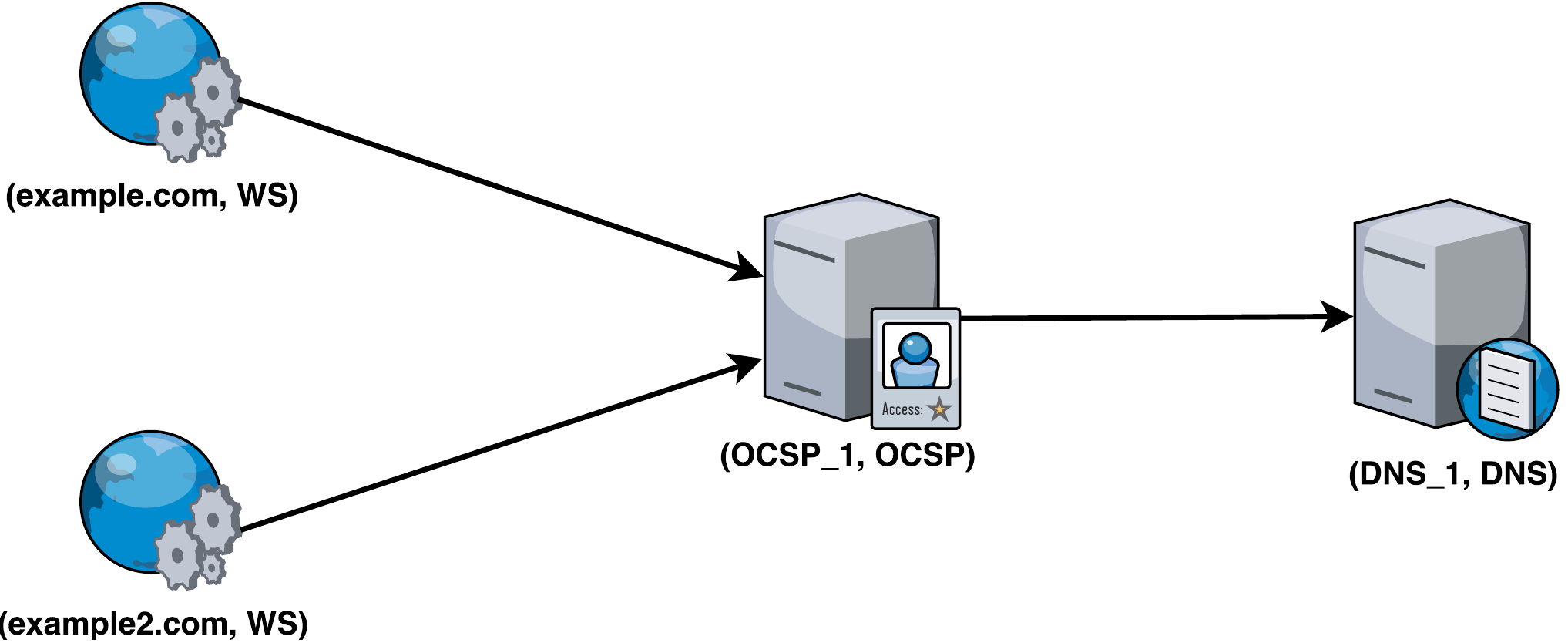}
\caption{An example of a transitive dependency} 
\label{fig:trans}
\end{figure}

In our analysis, we consider  two types of dependencies:
\begin{enumerate} 
\item {\em Direct Dependency:}
We say that a service $s_1$ has a direct dependency on service $s_2$ if $s_1$ uses $s_2$, It is denoted as $\orderedlist{s_1, s_2, \name{direct}}$.
 
From a robustness point of view, we are also specifically interested in the notion of {\em exclusive dependency}. That is, 
if service $s_1$ has a direct dependency on service $s_2$ where $s_2$ is of type \emph{A} and there exists no service $s_3$ of the same type \emph{A} such that $s_1$ has direct dependency on $s_3$.  For example, Figure \ref{fig:ddep} shows that \emph{(example.com,website)} is directly dependent on both $(DNS\_1, DNS)$ and $(DNS\_2, DNS)$ for service type \emph{DNS}.  However, it is exclusively dependent on $CDN\_A$ for service type \emph{CDN}.

\begin{figure}[!h]
\centering
\includegraphics[width=1\columnwidth]{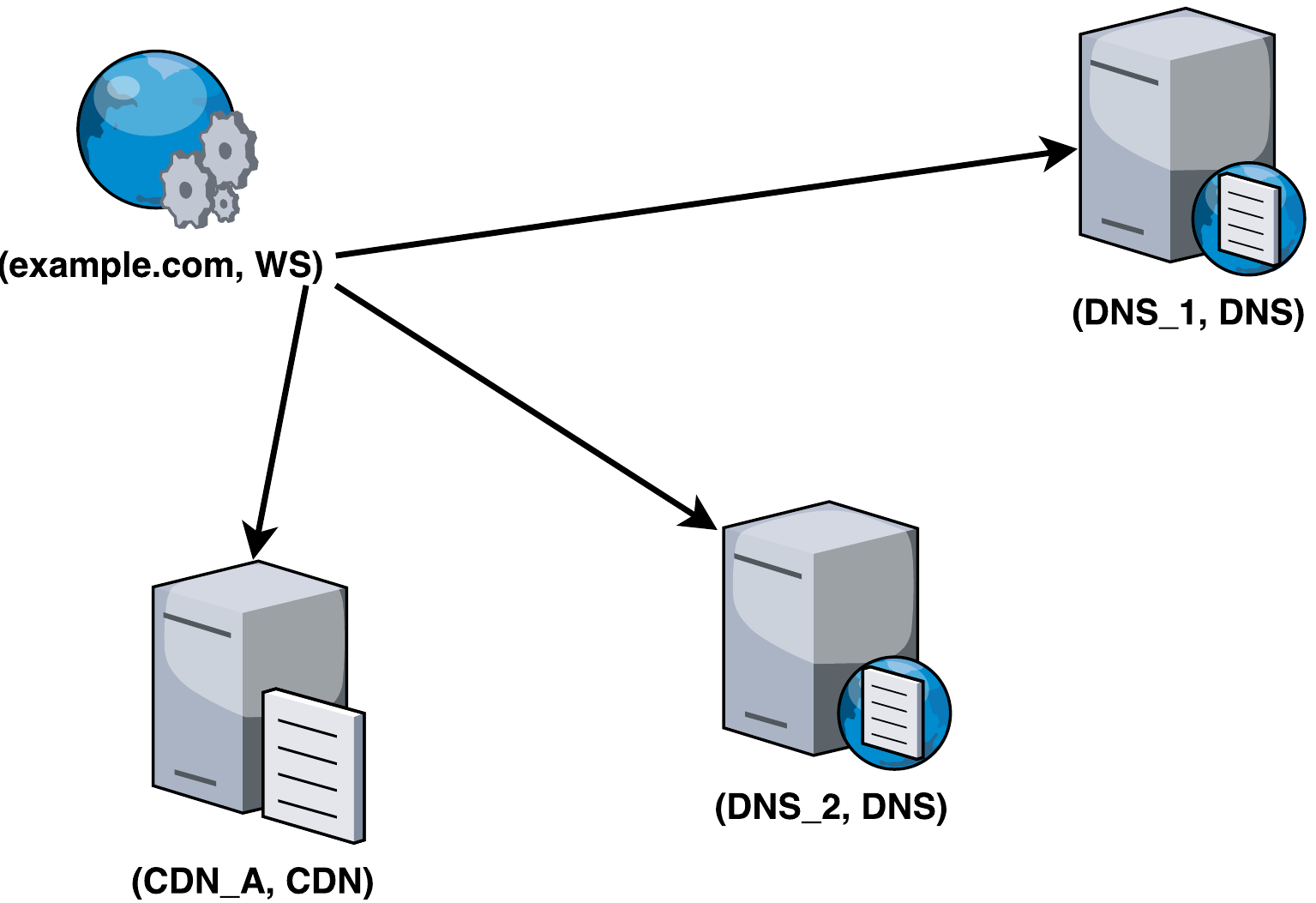}
\caption{The figure shows that \emph{example.com} has a direct dependency on DNS\_1 and DNS\_2 and exclusive dependency on CDN\_A.} 
\label{fig:ddep}
\end{figure}

\item{\em Indirect or Transitive Dependency:}\footnote{We use the terms indirect and transitive interchangeably to refer to this type of dependency throughout the paper.}

If a service $s_1$ has an exclusive dependency on service $s_2$ and $s_2$ has an exclusively dependency on service $s_3$, then we say that $s_1$ has a transitive dependency on service $s_3$.

Figure \ref{fig:trans} shows an example of transitive dependencies. If $DNS\_1$ fails, this results in the failure of $OCSP_1$ and all the websites exclusively dependent on it are affected as well.
\end{enumerate}

\subsection{Measurement Methodology}
To assess the hidden dependencies among \Webservices and third party infrastructure providers, we measure the authoritative nameservers (NS), CDNs (if any) and Certificate revocation information (OCSP servers and CRL distribution points) for Alexa's top 100,000 \Webservices. Alexa rank is calculated based on the traffic data which includes unique visitors and page views of a particular \Webservice \cite{alexarank}. The traffic across ranks drops roughly exponentially as shown by \cite{alexatraffic}\cite{rank2traffic}.
Therefore, we assume, top 100,000 \Webservices cover a significant amount of traffic on the Internet. Since Alexa's rankings change over time, we used a consistent list for all our measurements.
\\
In order to find out the authoritative name servers, we use dig (Domain Information Groper) which is a command line tool to query DNS servers \cite{dig}. We issue normal dig queries for the NS (nameserver) records which give the records for authoritative nameservers of these domains. We use Google's global DNS server \cite{googleDNS} to query this information. 
In order to distinguish between private and \Thirdpartyservice DNS providers, we used several indicators such as available lists of Third party DNS providers \cite{dnslist}, the number of \Webservices using a particular DNS provider. However, there is no accurate way to detect if a \Webservice is using a third party DNS provider. We observed 21419 DNS providers in total, 92\% of these serve less than 5 \Webservices in the top 100,000. We do not say anything about these DNS providers and since our analysis concerns the ones that predominantly serve \Webservices, we looked at all those DNS providers which serve more than 100 \Webservices. This list included total of 82 DNS providers, which we manually checked to see how many of these provide DNS services and found that 79 of them are Third party DNS providers. These 79 DNS providers cover 58\% of the total \Webservices. Therefore, for our analysis, we only consider these 79 DNS providers to be \Thirdpartyservices.  So our analysis might be an under-estimation.

Similarly the information related to OCSP servers and CRL distribution points can be extracted from the certificate of the respective \Webservice. To fetch certificates, we first send a SYN on TCP port 43 to see if the \Webservice supports HTTPS. If we receive a Connection Refused error, then it means the \Webservice does not support HTTPS. Next we initiate an HTTPS connection with it and fetch the certificates. We observed 69 distinct OCSP server and manually checked that 57 of them were \Thirdpartyservice. We detected this by looking up how many of these provide CA services to other \Webservices.

CDN information of a website can be extracted from its canonical (CNAME) records. When a \Webservice uses a CDN, the CNAME that it uses points towards the CDNs. Moreover, some CDNs set custom HTTP headers in the web request. To find the CDNs used by a website, we performed dig CNAME (canonical name) queries on all the links that belong to the same host, in a domain webpage. By looking at these records, we extract the CDNs using some indicators in the CNAME. We also look at HTTP headers to see for any custom headers set by the CDNs. A complete list of these indicators is mentioned here \cite{cdnlist}. However, we are not able to detect any CDNs that are not mentioned in this list. We observed 47 distinct CDNs and found that 44 of them are \Thirdpartyservice.

\begin{table}[thb]
\centering
\begin{tabular}{p{1cm}p{4cm}p{1cm}}
\hline
DNS & Total Unique DNS Providers & 21419 \\
\hline
\multirow{2}{*}{CDN} & Total \Webservices that use CDN & 29273  \\    & Total unique CDNs observed & 47   \\
\hline
\multirow{ 2}{*}{OCSP} & Total \Webservices that support HTTPS & 49170  \\    & Total unique OCSPs observed
 &  69  \\ \hline
\end{tabular}
\caption{Summary of the dataset collected for analysis}
\label{table2}
\end{table}

\subsection{Limitations}
We acknowledge some key limitations of the dataset and measurements to help put our results in perspective:

\begin{packeditemize}

\item First, we only study the interaction of a client with  DNS, CA,  CDN and other third-party providers for a particular \Webservice that are {\em externally visible}. For instance, we will miss dependencies if the interactions  are via APIs between the services that are   out-of-band; e.g., an ad exchange being contacted while the ad-service is being loaded.

\item Second, all of our measurements are conducted from a single vantage point in the US East coast. While we believe this is representative view of the structural dependencies of typical and popular \Webservices, we might miss region-specific dependencies; e.g., if the content rendered say in Asia-Pacific has a different dependency structure. 

\item Third, we do not measure the dependencies or robustness with respect to the physical and network  infrastructure; e.g., with respect to the physical hosting, or network routing, or physical capacity limitations. Thus, when we present our results as a service being potentially vulnerable it is a capacity- and hosting-agnostic analysis. Such data is proprietary and hard to obtain in practice; we believe that our analysis has value even with this limitation.  In particular, as prior attacks have shown it is not unreasonable that the adversary can suitably map the infrastructure and scale their capacity as needed~\cite{schneierblog,kang2013crossfire}.

\item Fourth, we do not consider dependencies that might arise from software vulnerabilities; e.g., if the services are running vulnerable versions of web/database servers or   content management systems.

\item Fifth, we do not focus on dependencies between webservices themselves; e.g., if some third-party widget or scripts are loaded. While this too can have implications for security (e.g., privacy), our focus in this paper is largely on the infrastructure 
 components like DNS, CDN, and CAs. We refer the readers to other related work for analysis 
  of third-party web content~\cite{roesner2012detecting,butkiewicz2011understanding}.

\item Finally, we only analyze the dependencies  as observed on the  landing pages of the popular websites. While we believe this is representative of common workload patterns, we do acknowledge that we do not measure  dependencies  that may manifest deeper in the content hierarchy. 

\end{packeditemize}

%\section{Data Collection}
%\subfile{data_collection}

%\section{Analysis}
% \subfile{analysis}

%TO DO:

\section{Analyzing Direct Dependencies}
\label{sec:direct}

In this section, we consider the direct dependencies that \Webservices have on various \Thirdpartyservices. We analyze how robust individual \Webservices are in using \Thirdpartyservices and also find that a small number of \Thirdpartyservices are critical for a large fraction of \Webservices.  

\subsection{Robustness of \Webservices}

\begin{insight}
 Third-party services and content are widely used --76.7\% \Webservices  use one or more \Thirdpartyservice  infrastructure providers and 90.7\% of \Webservice entail loading third-party content from other \Webservices
\label{insight:exclusive}
\end{insight}

Table~\ref{table:thirdpartyprevalence} shows that out of the top-100K sites, more than 90 \% use at least one \Thirdpartyservice for their operation. Several studies have shown that modern webservices load content from multiple third-parties~\cite{roesner2012detecting,butkiewicz2011understanding}. Most of these prior studies have focused on the privacy leakage implications of third-party content and on potential performance bottlenecks in the web browser~\cite{butkiewicz2015klotski}. Our focus here is on the robustness and dependencies on third-party services. 

We further break these dependencies into common infrastructure services (e.g., DNS, CDN, and OCSP) vs. third-party \Webservices (e.g., for ads or analytics or other widgets). We find that almost 77\% of sites use at least one third-party infrastructure service provider (i.e., CDN, DNS, or OCSP). This confirms that third-party service providers are indeed a fact-of-life for modern webservices.  

\begin{table}[!ht]
\centering
\begin{tabular}{ |p{3cm}|p{2cm}| }
\hline
\Thirdpartyservices & Percentage of \Webservices   \\
\hline
Ad services & 62.1 \\
Analytics & 55 \\
Overall Third-party resources & 90.7 \\

Third-party infrastructure & 76.7  \\
\hline
\end{tabular}
\caption{Prevalence of \Thirdpartyservices in the top 100k \Webservices}
\label{table:thirdpartyprevalence}
\end{table}

Next, we look in greater depth at the specific infrastructure services such as DNS, CDN, and OCSP as they can have key security implications for the correctness and availability of a given \Webservice.

\begin{insight}
Out of top 100k \Webservices, 58\% use \Thirdpartyservice DNS. 96.6\% of these \Webservices are fragile as they have an exclusive dependency on a Third party DNS provider. This constitutes 55.6\% of the top 100k \Webservices.
\label{insight:dns:exclusive}
\end{insight}

We begin by analyzing the robustness of 
 \Webservices in their use of \Thirdpartyservice for DNS. For instance, Twitter and Spotify both relied on Dyn for hosting authoritative name services. 

For DNS, robustness means that a \Webservice should not have exclusive dependency on any single DNS provider because in case of an attack or any other kind of failure, these will be affected as compared to the ones who use multiple providers. 

\begin{figure}[!t]
\centering
\includegraphics[width=1\columnwidth]{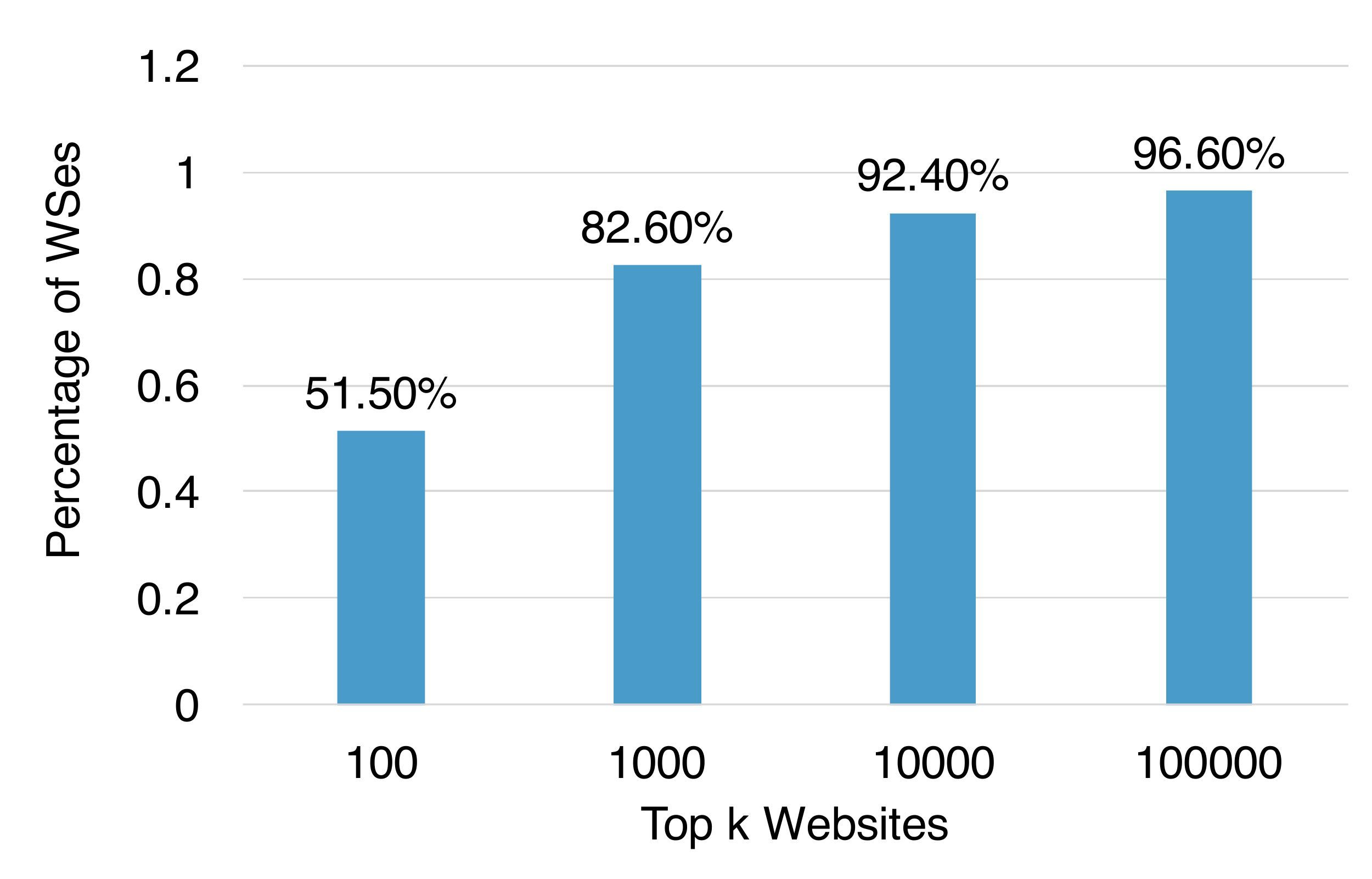}
\caption{The percentage of \Webservices that use a \Thirdpartyservice DNS provider and are fragile since they exclusively depend on one DNS provider. We plot data for different \Webservices rank ranges (top 100, top 100,000). More popular \Webservices (51.5\% top 100), are more robust than the less popular ones (96.60\%, top 100K).} 
\label{fig:robDNS}
\end{figure}

Figure~\ref{fig:robDNS} shows the fraction of \Webservices that have an exclusive dependency on a single \Thirdpartyservice DNS provider. We observed that out of the top 100K sites, 58\% \Webservices used third party DNS providers. 96.6\% of them are not robust and have an exclusive dependency on one \Thirdpartyservice DNS. We also observe that this percentage fragility of \Webservices varies across the different popularity rank ranges.  \Webservices that are more popular are marginally more likely to be robust  compared to the ones that are less popular. For instance, in the top 100, 33 \Webservices in total use third party DNS and  50\% of them are fragile. 
 
 This observation shows that  most \Webservices today do not use  \Thirdpartyservices in a robust manner. We also looked at the Dyn incident to analyze the reaction of \Webservices to the whole incident. Table \ref{table:dynStats} shows the number of \Webservices that used Dyn before and after the impact. It can be seen that the prevalent behavior among \Webservices was to not change anything. However, we see some cases where \Webservices became robust by having a secondary DNS provider. We also found that the \Webservices that left, 73\% of these went to other \Thirdpartyservices and only 5 of these left Dyn and joined two other \Thirdpartyservices DNS providers. A total of 5 \Webservices that left shifted to having their own private DNS among the top 10k \Webservices.
 We revisit this in Section~\ref{sec:discuss} when we elaborate on recommendations for different stakeholders.

% Figure

\begin{table}[!ht]
\centering
\begin{tabular}{ |p{3cm}|p{1cm}|p{1cm}|p{1cm}| }
\hline
& 100 & 1000 & 10,000 \\
\hline
Total No. of \Webservices before & 19 & 103 & 483 \\
No. of robust \Webservices before & 11 & 41 & 151 \\
No. of \Webservices affected & 8 & 62 & 332 \\
\Webservices that left & 3 & 10 & 50 \\
\Webservices that left and shifted to private & 0 & 1 & 4 \\
\Webservices that became robust & 3 & 21 &74 \\
\Webservices that did nothing & 2 & 31 & 208\\
\hline
\end{tabular}
\caption{The table shows how \Webservices reacted to the Dyn incident in the top 100, 1000 and 10,000 \Webservices}
\label{table:dynStats}
\end{table}

\begin{insight}
Out of the 49.2\%  \Webservices which support HTTPS,    96\%  use \Thirdpartyservice OCSPs and roughly 20\% use OCSP stapling to lessen this dependency.
\label{insight:ocsp:exclusive}
\end{insight}

An important component of the HTTPS workflow is the use of OCSP or CRL servers to check the validity 
 of the certificate. As observed elsewhere~\cite{stark2012case,istlsfast}, this step can add non-trivial latency to the TLS handshakes and there are several ongoing efforts to speed this protocol~\cite{stark2012case}.
 
 Out of the set of top 100k \Webservices we analyzed, roughly half (49.2\%) support HTTPS. These numbers are consistent with recent reports on the increased adoption of HTTPS. Of this set of \Webservices that are HTTPS-enabled, almost all of them (96\%) rely on \Thirdpartyservices for OCSP.
 
Now, one well-known approach for a \Webservice to be robust and remove this third party dependency is 
 to use OCSP Stapling~\cite{RFC6961}. 
  OCSP Stapling effectively  allows the \Webservice to bear the resource cost involved in providing OCSP responses by appending a time-stamped OCSP response signed by the CA to the initial TLS handshake, thus eliminating the need for clients to contact the CA.
  
\begin{figure}[!t]
\centering
\includegraphics[width=1\columnwidth]{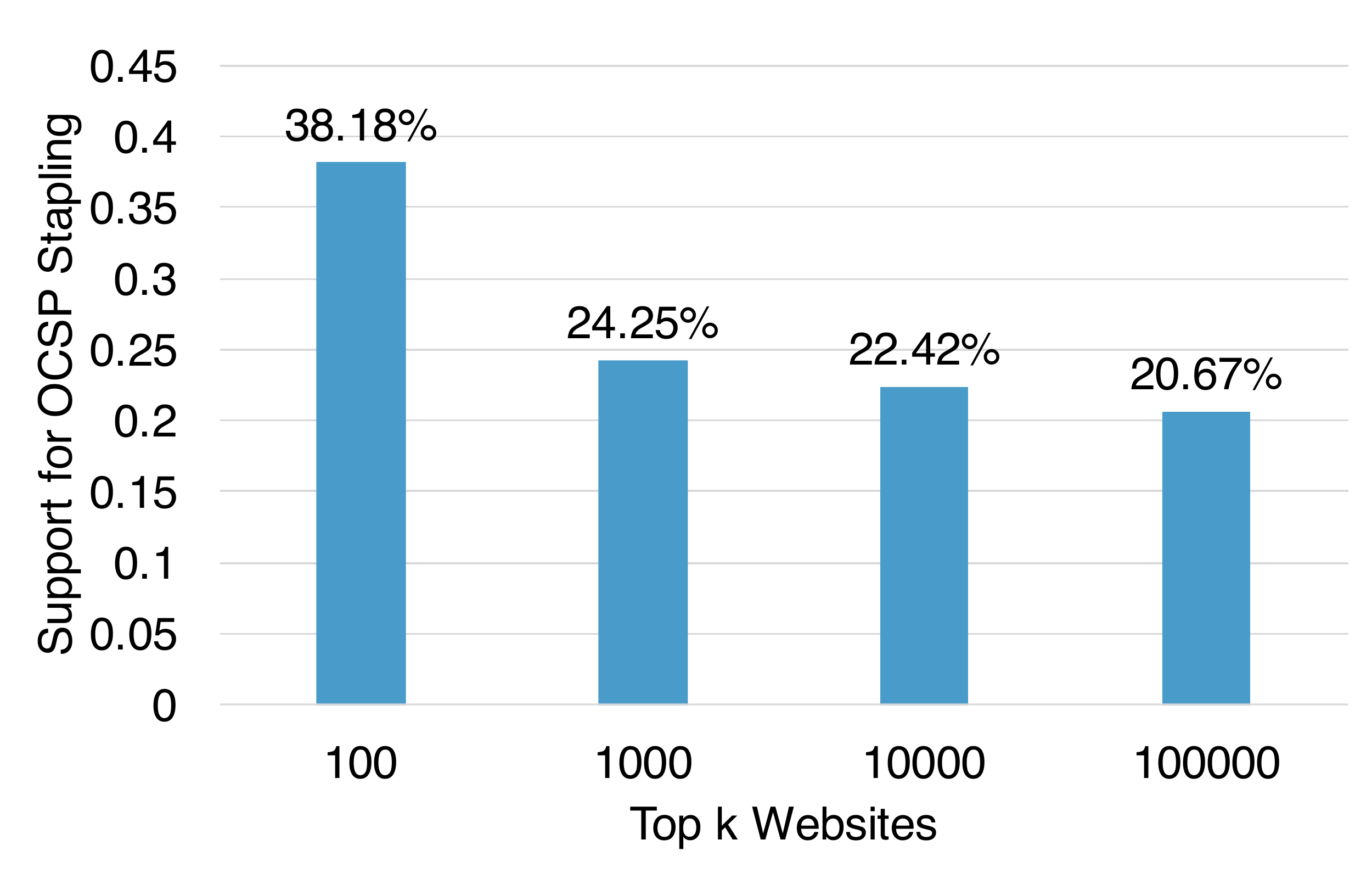}
\caption{The percentage of \Webservices that support OCSP Stapling. We plot data for different \Webservices rank ranges (top 100, top 100,000). Support for OCSP Stapling varies across ranks with more popular \Webservices (38.18\%, top 100) showing higher \%age of OCSP Stapling support as compared to less popular ones (20.67\%, top 100K)} 
\label{fig:ocspRobust}
\end{figure}

 However, unlike OCSP and CRLs which involve the CA, to support OCSP \Webservice administrators have to enable it. Unfortunately, as we see in Figure~\ref{fig:ocspRobust}, only a small fraction of the \Webservices use OCSP Stapling and are thus  exposed to potential RoQ and RoS attacks.

% ----------------------------------------
% CDN
%\paragraph{$\Webservices \rightarrow CDNs$}
\begin{insight}
30\% of \Webservices use CDNs, 96.7\% of these use \Thirdpartyservice CDNs, and  93\% of these, have an exclusive dependency on one CDN provider. 
\label{insight:cdn:exclusive}
\end{insight}

Robustness in case of a CDN is similar to that of a DNS provider. A \Webservice should ideally not have an exclusive dependency on one particular CDN.  
While one CDN might offer 100\% uptime, choosing more than one CDN means having a fallback in case of an outage because traffic will get rerouted to the other CDN. 

\begin{figure}[!t]
\centering
\includegraphics[width=1\columnwidth]{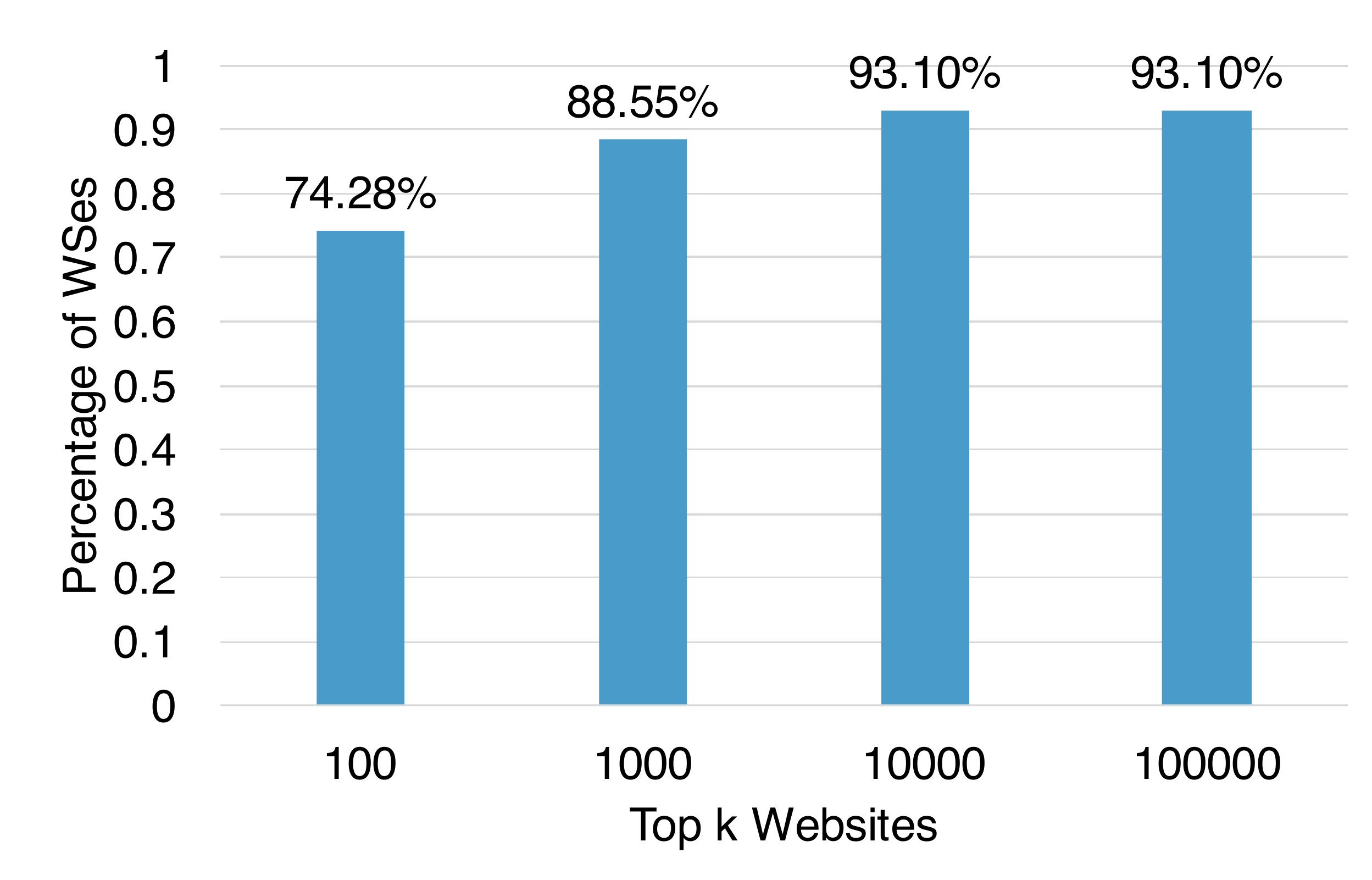}
\caption{The percentage of \Webservices that use a \Thirdpartyservice CDN provider and are fragile since they exclusively depend on one CDN provider. We plot data for different \Webservices rank ranges (top 100, top 100,000). More popular \Webservices (74.28\% top 100), are comparatively more robust than the less popular ones (93.10\%, top 100K).} 
\label{fig:cdnR}
\end{figure}

We analyzed the  top 100,000 \Webservices, and observe that roughly 30\% of these use some  form of CDN, either private (E.g., large providers like Google or Netflix often run their own CDN infrastructure) or  \Thirdpartyservice CDN services. 
 The vast majority of \Webservices (96.7\%) that use CDNs rely on \Thirdpartyservices, which is unsurprising given the massive infrastructure and management costs associated with operating CDNs. Of these, we find that 93\% of them are not robust as shown in figure \ref{fig:cdnR}. Similar to DNS, this degree of robustness varies across ranks with more popular \Webservices being more robust.

\paragraph{Implications}
We observe that when we combine the sets of \Webservices that are exclusively dependent on either \Thirdpartyservice CDN, OCSP or DNS provider, we get 73.14\% of  \Webservices that have a single point of failure. 
The above observations on the exclusive dependencies on \Thirdpartyservices has key implications for the \RoQ, \RoS, and \DoS attacks we considered in Section~\ref{sec:background}.
For instance, the fragility in $\Webservice \rightarrow DNS$ dependency has key implications for \DoS and \RoQ for the users as we have seen with the Dyn incident, and 55.6\% of \Webservices are vulnerable to this. The non-robust use of 
  CAs, $\Webservice \rightarrow OCSP$ leads to threats like \RoS and \RoQ as observed in prior work~\cite{stark2012case} and 85.2\% of \Webservices that support HTTPS are vulnerable to these kind of attacks. Finally, the exclusive dependency on a single third party CDN $\Webservice \rightarrow CDNs$ can leads to threats like \DoS and \RoQ and 93.3\% of \Webservices that use CDNs are vulnerable to this.
  In general, we find that a majority (73.14\%) of the top 100k \Webservices are prone to attacks. In such a scenario, a common dependency among these fragile \Webservices can cause huge damage.

\subsection{Concentration in Third-party services}
% ----------------------------------------
%  OBSERVATION 2
The previous analysis shows that a large fraction of \Webservices are not robust in their use of \Thirdpartyservices. Now, what it does not show is whether a large fraction of these \Webservices rely on the same subset of \Thirdpartyservices or if there is a substantial diversity in this pattern. In this section, 
 we focus on the concentration behavior where a small 
  number of  \Thirdpartyservices predominantly offer these services. This type of concentration and the   clustering  does not bode well as these \Thirdpartyservices become  good targets for attack that can effectively impact a large number of \Webservices (e.g., Dyn attack took down a large number of popular services).

\begin{figure}[!t]
\centering
\includegraphics[width=1\columnwidth]{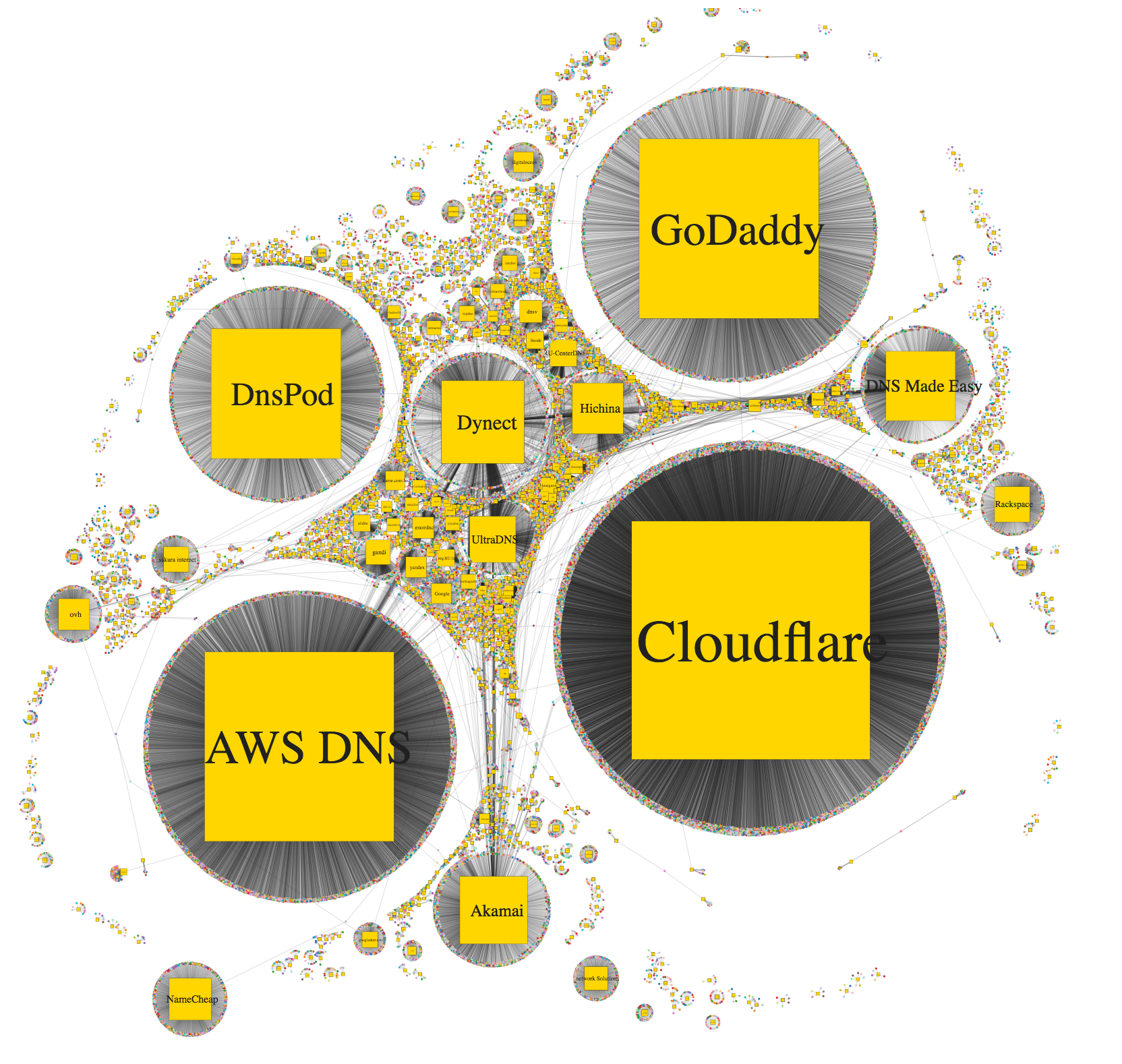}
\caption{The figure shows dependency graph of top 100k \Webservices and their DNS providers. The size of DNS provider node is proportional to the number of websites using them. It can be seen that Cloudflare is the major provider for the top 100k \Webservices} 
\label{fig:dnsG}
\end{figure}

\begin{insight}
Top 10 \Thirdpartyservice DNS providers cover 37.8\% of the top 100k \Webservices.
\label{insight:concentration:dns}
\end{insight}

\begin{figure}[!t]
\centering
\includegraphics[width=1\columnwidth]{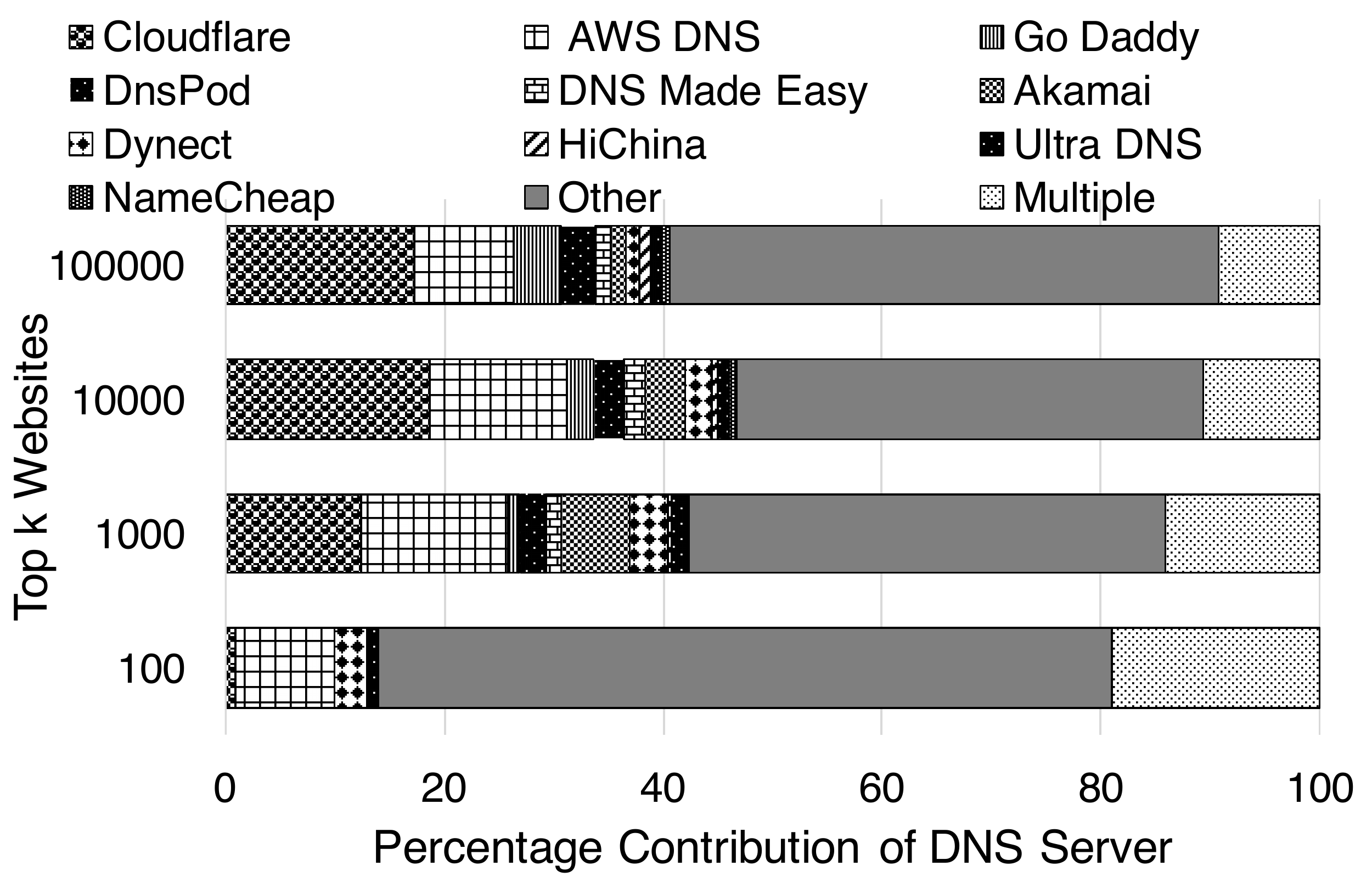}
\caption{The figure shows percentage contribution of top 10 DNS providers for different rank ranges (100,100k). The graph only includes exclusive dependency on a particular provider. Websites that use multiple providers are shown as``Multiple''. It can be seen that top 10 DNS providers constitute almost 38\% of the total \Webservices} 
\label{fig:dns}
\end{figure}

Figure~\ref{fig:dns} shows the fraction of  \Webservices covered by the top-K DNS for different \Webservice popularity ranges. For clarity of presentation, the figure only considers the coverage by analyzing the exclusive dependences of \Webservices on a particular provider. (The \Webservices that have more than one DNS providers are marked as ``multiple'' in the figure.)   In total, we observed 21,419 DNS providers in total for our set of \Webservices. We do see a long-tailed behavior, where  19,650 served less than 5 \Webservices. However, among the remaining 1769 providers, we see that a small number of providers cover almost 50\% of the \Webservices as seen in   Figure~\ref{fig:dns}.

Among  these \Thirdpartyservice DNS providers, we see that Cloudflare is the major provider and covers almost 17\% of the total 100k \Webservices. However, within the top 100 websites, AWS DNS which is Amazon route 53 DNS service and Dyn are the dominant ones. Furthermore, the contribution of AWS DNS is almost the same across all ranks of \Webservices. This ranking might be important because more popular \Webservices carry more traffic and hence can do more damage.

\begin{figure}[!t]
\centering
\includegraphics[width=1\columnwidth]{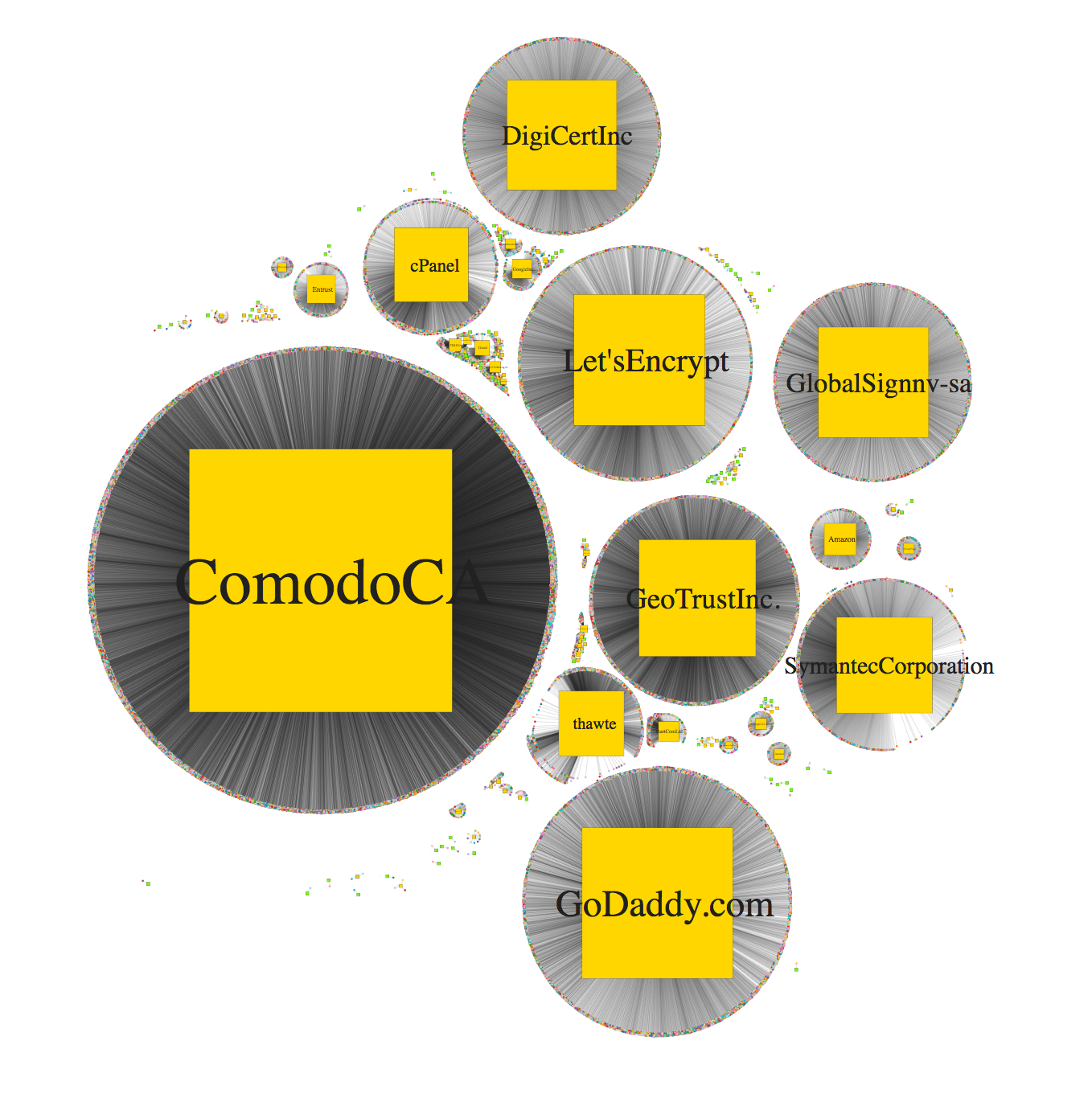}
\caption{The figure shows dependency graph of top 100,000 websites and their OCSP providers. The size of OCSP provider node is proportional to the number of websites using them.It can be seen that Comodo is the major CA among all 100k \Webservices.} 
\label{fig:ocspG}
\end{figure}
\begin{figure}[!h]
\centering
\includegraphics[width=1\columnwidth]{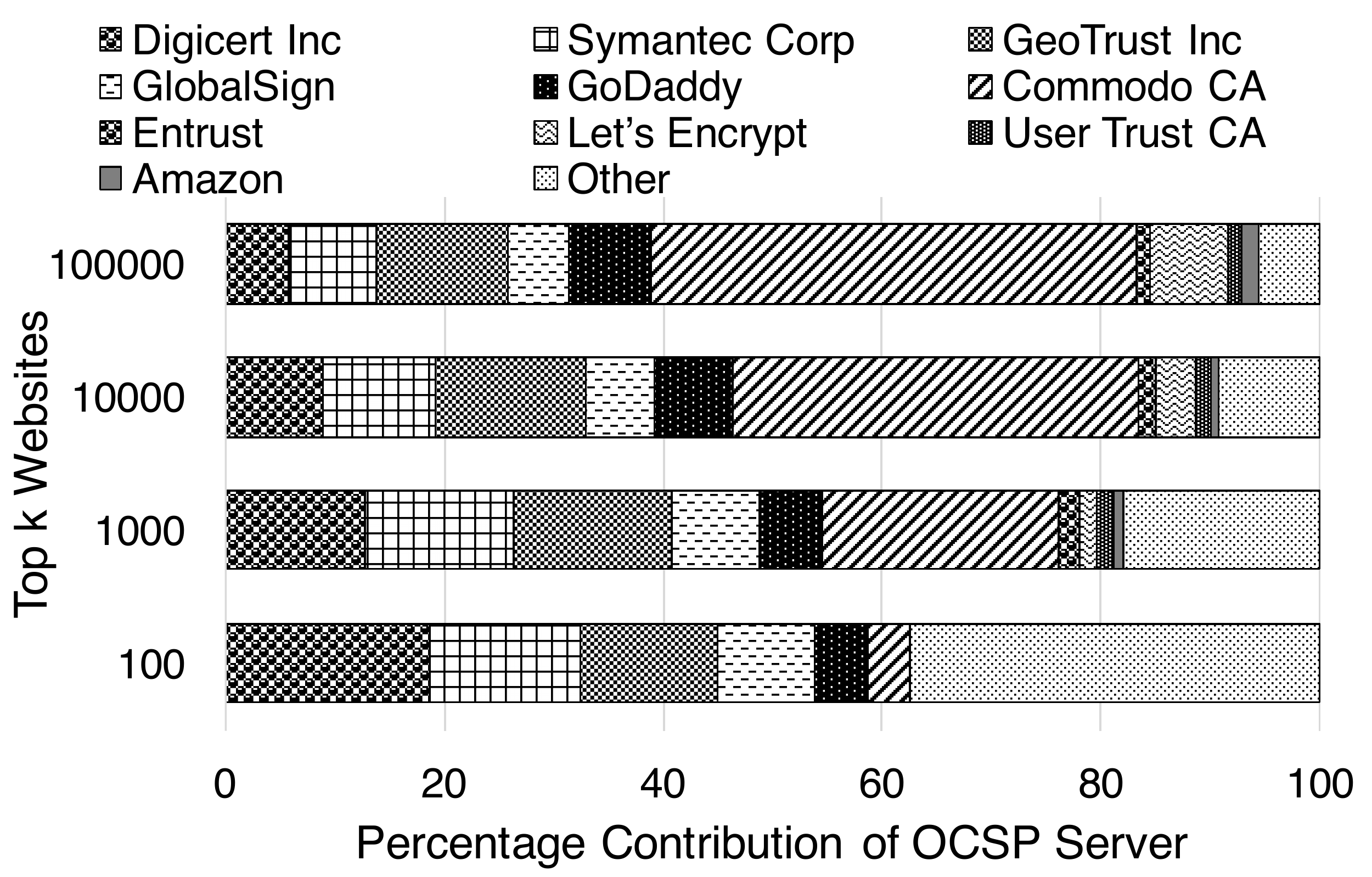}
\caption{The figure shows percentage contribution of top 10 CAs that host OCSP servers for \Webservices in various rank ranges (100,100k). It can be seen that top 10 major providers cover almost 95\% of the total HTTPS-enabled \Webservices} 
\label{fig:ocsp}
\end{figure}

%\subsection{$\Webservices \rightarrow CAs$}
\begin{insight}
Top 10 \Thirdpartyservice CA cover 94.3\% of the \Webservices that support HTTPS. This constitutes 46\% of the top 100k \Webservices.
\label{insight:concentration:ocsp}
\end{insight}
Figure \ref{fig:ocsp} shows the percentage distribution of CAs that host OCSP servers for top k websites. In our measurements, we found that the $\Webservice \rightarrow CA$ dependency is always exclusive. In total, we observed 69 unique OCSP servers for our set of \Webservices. However, from the figure we can see that out of these, just 7 services
 cover more than 90\% of the \Webservices across each rank range. Among these 7 services, Symantec, GeoTrust, Digicert are the dominant providers across all rank ranges. Interestingly, Comodo  which is the most popular service  in the overall set, only serves a small number of \Webservices in top 100. This behavior is similar to Cloudflare in DNS providers. Investigating this further, we found that most of these ( 88\% ) of these \Webservices actually use Cloudflare as their DNS and CDN, which suggests a hidden correlation as well.

\begin{insight}
Top 10 \Thirdpartyservice CDN providers cover 86.4\% of the \Webservices that use CDNs. This constitutes 25.3\% of the top 100k \Webservices. 
\label{insight:concentration:cdn}
\end{insight}

\begin{figure}[!h]
\centering
\includegraphics[width=1\columnwidth]{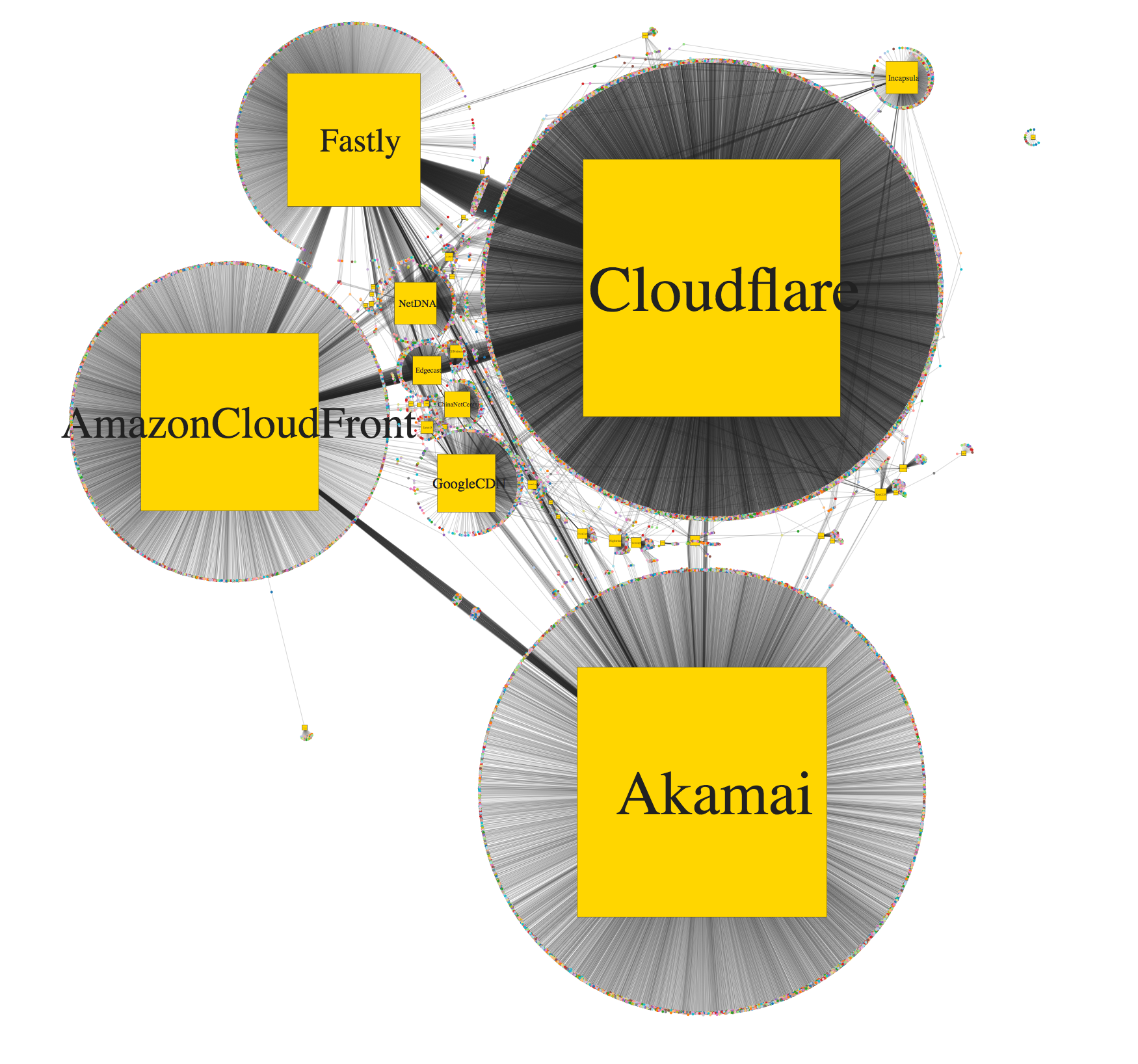}
\caption{The figure shows dependency graph of top 100,000 websites and CDN providers. The size of CDN node is proportional to the number of websites using them. It can be seen that Cloudflare and Akamai are among the two major providers.} 
\label{fig:cdnG}
\end{figure}
\begin{figure}[!ht]
\centering
\includegraphics[width=1\columnwidth]{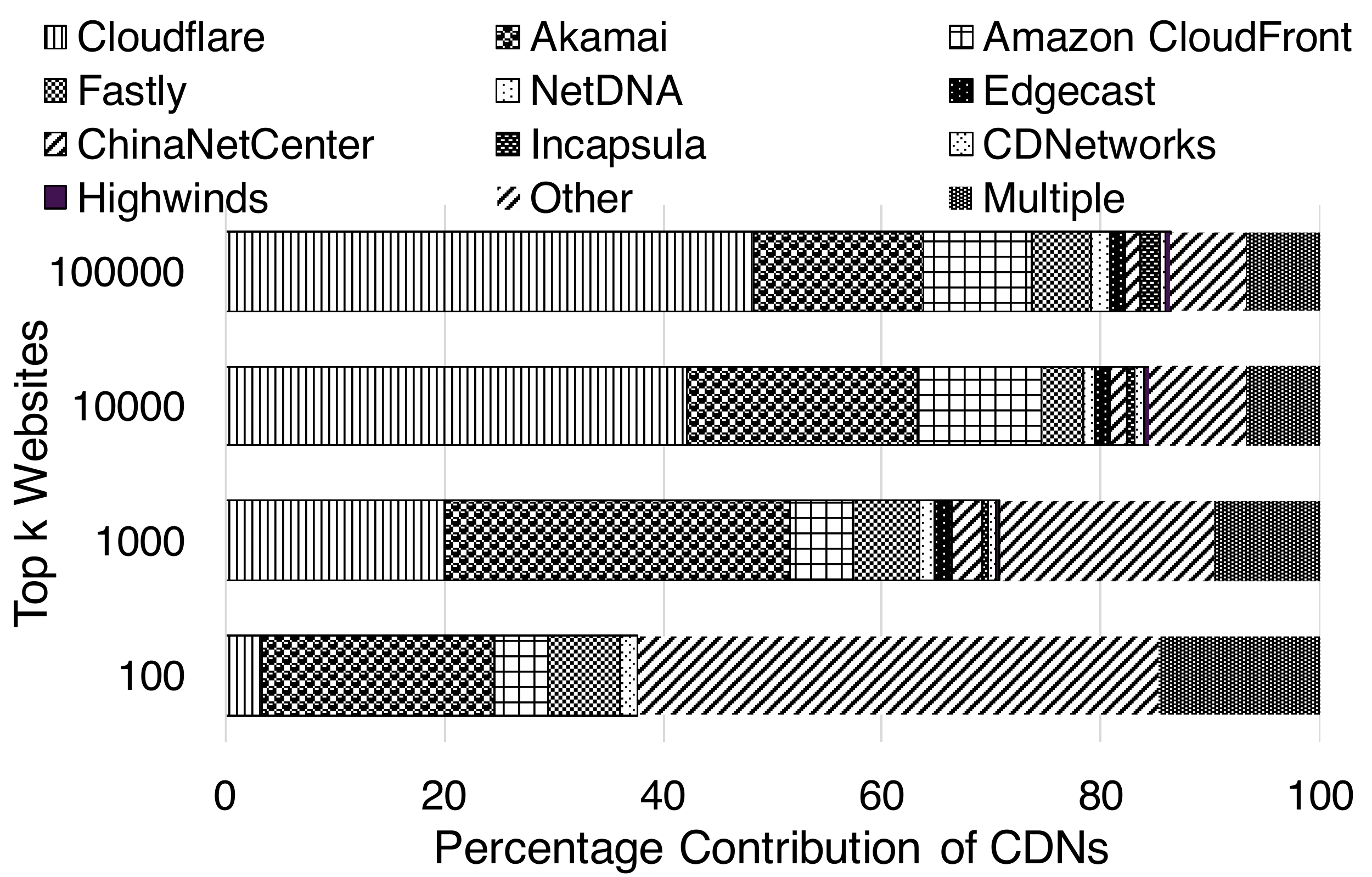}
\caption{The figure shows percentage contribution of top 10 \Thirdpartyservice CDNS for \Webservices in various rank ranges (100,100k). It can be seen that top 10 major providers cover almost 86\% of the total \Webservices that use CDNs} 
\label{fig:cdn}
\end{figure}
We observed 47 \Thirdpartyservice CDNs in total for the top 100,000 \Webservices. If we look at figure \ref{fig:cdn}, it is quite evident that there are a few major providers like Cloudflare, Akamai, Amazon Cloudfront, that are dominant across all ranks. Cloudflare, here too, is not dominant in top 100 \Webservices. Figure \ref{fig:cdnG} gives a view of these clusters and the existence of a few major providers.

\paragraph{Implications} At first glance, the web ecosystem may appear   decentralized with a low barrier of entry. However, the reality is that  a handful of \Thirdpartyservice providers serve most of the popular \Webservices. As we observed, in case DNS, top 10 providers cover 37.8\% of the top 100k \Webservices. Similarly top 10 CDN providers cover 25.3\% of them while top 10 OCSP servers cover 46\% of the total 100k \Webservices. This situation is a bit alarming, considering almost 50\% of the top 100k \Webservices being managed by just 10 providers. Consequently, if these providers are under attack or there is some kind of misconfiguration at their  end, then a huge part of the web ecosystem will likely be unavailable or function incorrectly.  In fact, these are the high profile targets that attackers (e.g., the Mirai botnet) may consider  (and likely already do) as potential targets for future attack campaigns.The top 10 CDNs and DNS might be very food targets for RoQ and DoS attacks. Similarly, top 10 OCSP providers, will be a very good target for RoS and RoQ attacks, particularly if the aim of the attacker is to maximize the total damage caused.

\section{Analyzing Transitive  Dependencies}
\label{sec:indirect}

During the Dyn incident, many \Webservices went down. 
 For many of these, the downtime  was directly attributed to Dyn; i.e., the ones that had a \textit{direct} dependency on Dyn.  A perhaps less prominent observation was that a few key CDNs, such as Fastly, also were unreachable and consequently any \Webservice that used Fastly was therefore also down. The reason for this was that Fastly was using Dyn, and thus there was an \textit{indirect} dependency of \Webservices that used Fastly, on Dyn \cite{fastly}.

Motivated by this observation, we now systematically study the extent of  these indirect, potentially ``hidden'', and transitive dependencies. To do so, we again focus on the three key \Thirdpartyservices: DNS, CDN, and OCSP, and determine how many of the popular services from each of these categories are dependent on each other (e.g. $OCSP \rightarrow DNS$, $OCSP \rightarrow CDN$, etc). We then do a what-if analysis on the potential impact of these indirect dependencies and which \Thirdpartyservices are most critical in case of an attack. We also study how the landscape changes when we consider both indirect and direct dependencies, as compared to direct dependencies alone.

We consider three different kinds of intra-\Thirdpartyservice dependencies: $OCSP \rightarrow CDN$, $OCSP \rightarrow DNS$ and $CDN \rightarrow DNS$.\footnote{The other dependencies do not  make sense in practice and naturally do not manifest in the real world.}
To get this data, we use the same approach that we used for \Webservices. To find the DNS used by CDNs, we look at the custom CNAMEs set by each CDN for various \Webservices and from that we extract the CDN domain to perform dig queries on it. Similarly to find the DNS of OCSP servers, we extract the OCSP URL from the certificate and perform dig queries on the host. Furthermore, for finding CDNs used by OCSP servers, we check CNAME for the hosts extracted from the OCSP URL and use the same approach as mentioned in Section \ref{sec:methodology}. We now consider each of them separately. Note that as before, for each of the indirect dependencies, we mark the \Thirdpartyservice that another \Thirdpartyservice depends on as not robust if it is an exclusive dependency.  
 
\subsection{Robustness of \Thirdpartyservices}

\begin{table}[th]
\centering
\begin{tabular}{|p{2.1cm}|p{2cm}|p{1cm}|}
\hline
Dependency & Total \Thirdpartyservice Dependencies & Fragile \\
\hline
$CDN \rightarrow DNS$  & 19 & 16\\
$OCSP\rightarrow DNS$  & 34 & 29\\
$OCSP\rightarrow CDN$  & 18 & 18\\
\hline
\end{tabular}
\caption{The table shows for each dependency, the number of provider dependent on \Thirdpartyservice provider and the number of fragile provider among them. }

\label{table:thirdpartyrobust}
\end{table}

\begin{insight}
40.4\% of CDNs and 49\% of OCSP services use a \Thirdpartyservice DNS provider.  85\% of these CDNs and OCSP services are not robust. Similarly, 33.3\% of OCSP providers use a \Thirdpartyservice CDN, and 100\% of these OCSP services are not robust. 
\label{insight:indirect_all}
\end{insight}

Table \ref{table:thirdpartyrobust} provides a summary of our results analyzing these intra-\Thirdpartyservice dependencies discussed above. 

$CDN \rightarrow DNS$ dependency:  CDNs typically have their own DNS infrastructure so that they have more control over where customer traffic gets routed to. However, we found that this was not quite the case. Out of a total 47 observed CDN providers, 19  (40.4\%) actually use a \Thirdpartyservice, of which 16 (84.2\%) are exclusively dependent on a single DNS provider. The two top DNS providers for CDNs are AWS-DNS and Dyn, with AWS DNS serving 9 CDNs.   (These CDNs do not provide services to a significant amount of \Webservices and are not among the major providers.)

$OCSP \rightarrow DNS$ dependency: Out of  69  OCSP providers, 34 of them (49\%) use \TPS DNS providers, of which 29 (85.2\%) were exclusively dependent on a single provider, and thus are not robust.   Interestingly, five of these OCSP providers that are not robust rank within the top-10 most used OCSP service providers and are used by 34\% of the \WSES that support HTTPS.

$OCSP \rightarrow CDN$ dependency: We observed that many CAs host OCSP servers on CDNs for performance gain. Out of the 69  OCSP providers, 18 of them (26.1\%) use \TPS CDN  providers, of which all 18 were exclusive dependencies (100\%). 5 of these are among the top 10 OCSP providers and cover 35.3\% of the total \Webservices that support HTTPS.

\paragraph{Implications} This is a surprising result that many of these \Thirdpartyservices are themselves interdependent and moreover have non-robust dependencies. This suggests that \Webservices who rely on these non-robust \Thirdpartyservices may need to take additional safeguards; e.g., either adding a layer of redundancy themselves or switching to other more robust \Thirdpartyservices.

\subsection{Impact of transitive dependencies}
We looked into the transitive dependencies induced by intra-dependency of \Thirdpartyservices. Specifically, we consider 
 four-types of transitive dependencies: (1) $\Webservices \rightarrow OCSP \rightarrow DNS$; (2)  $\Webservices \rightarrow OCSP \rightarrow CDN$; (3)  $\Webservices \rightarrow CDN \rightarrow DNS$; and (4)  $OCSP \rightarrow CDN \rightarrow DNS$. Next, we 
  characterize the impact of such transitive dependencies on the popular \WSES.

\begin{insight}
With indirect dependencies, 37.2\% of the top 100,000 \WSES are now dependent on just the top-10 CDN providers. Similarly, due to indirect dependencies 56.4\% of the top 100,000 \WSES are now dependent on the top-10 DNS providers.  
\label{insight:cdn:CoverageIndirect}
\end{insight}
Based on these transitive dependencies, next we characterize how the top 100,000 \WSES were affected in terms of their robustness. Due to indirect dependencies, 37.6\% of the top 100,000 \WSES are now dependent on the top-10 CDN providers. In contrast, when considering direct dependencies only, \textit{all} CDN providers collectively covered only 30\% of the top 100,000 \WSES. Note that now, in addition to the direct dependency, a \WS can be dependent on a particular CDN provider, via an OCSP.

Similarly, due to indirect dependencies 56.4\% of the top 100,000 \WSES are now dependent on the top-10 DNS providers. In comparison, 58\% of the top 100,000 websites were directly dependent on \textit{all} third party DNS providers. 
In other words, with transitive dependencies a significant portion of the most popular websites become dependent on relatively few DNS and CDN service providers, increasing the concentration effects we saw earlier in Section~\ref{sec:direct}.

\begin{figure}[!ht]
\centering
\includegraphics[width=1\columnwidth]{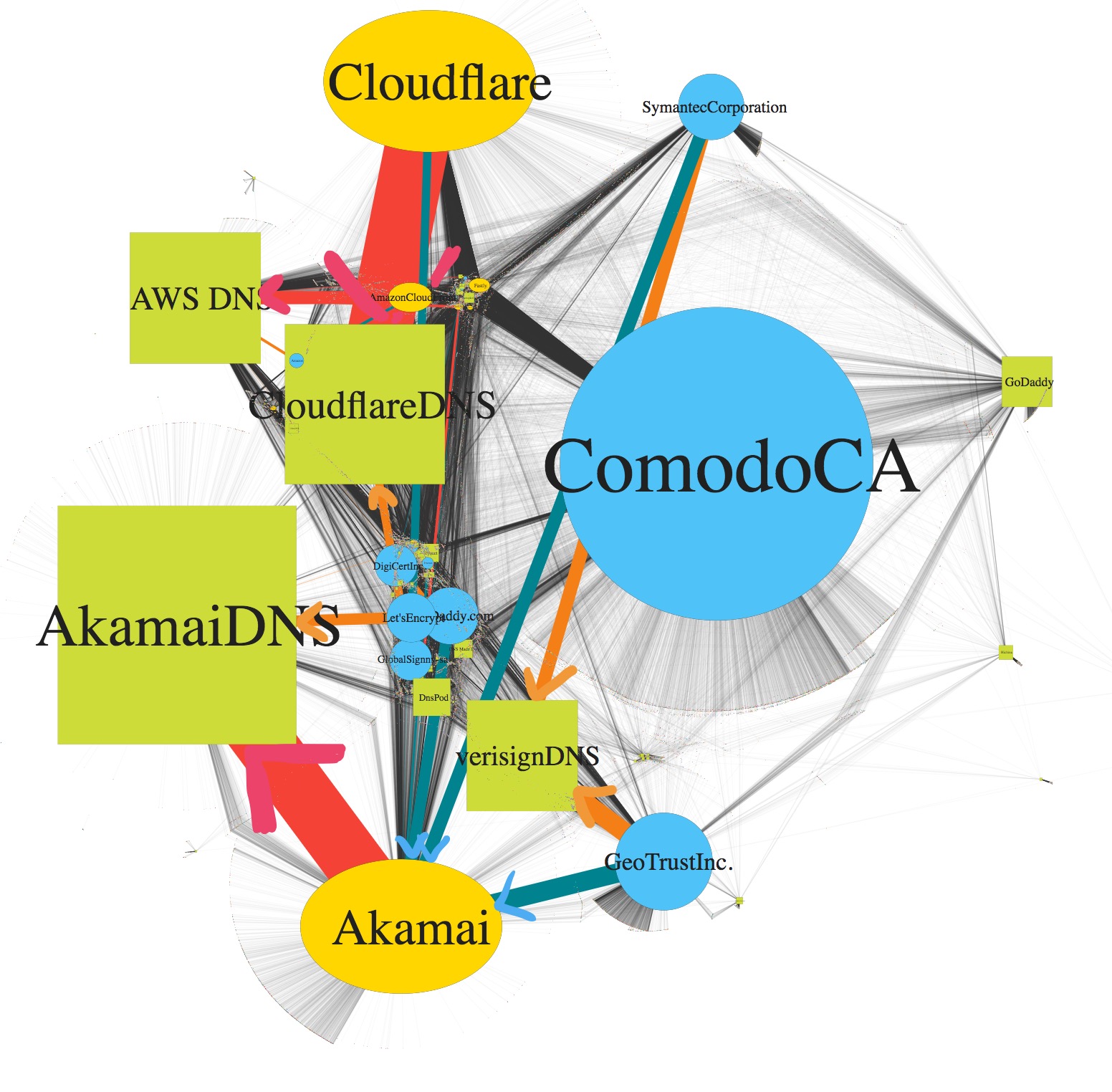}
\caption{The figure shows the overall dependency graph of top 100,000 \Webservices on various \Thirdpartyservices. The size of each node is proportional to their collateral damage. The weight of each edge is proportional to the collateral damage of the source node. The graph also shows intra-\Thirdpartyservice dependencies. It can be seen that Akamai DNS, Comodo CA etc are among the major \Thirdpartyservice providers.}
\label{fig:collateral}
\end{figure}
\begin{insight}
Transitive dependencies can amplify the total damage caused by a \Thirdpartyservice provider and the structure of the top 10 highest impact  \Thirdpartyservice DNS and CDNs changes when we consider transitive dependencies.
\label{insight:damage}
\end{insight}

Figure \ref{fig:collateral} represents a weighted graph showing the transitive relation among various \Thirdpartyservices and \Webservices, where the edge weights are proportional to the in-degree of the source node (thicker lines means higher weight). 
 The size of each node in the graph is proportional to the number of \Webservices dependent on them. We next consider each service DNS, CDN and CA and show that transitive dependencies amplify the collateral damage caused by each provider.

\begin{figure}[!t]
\centering
\includegraphics[width=1\columnwidth]{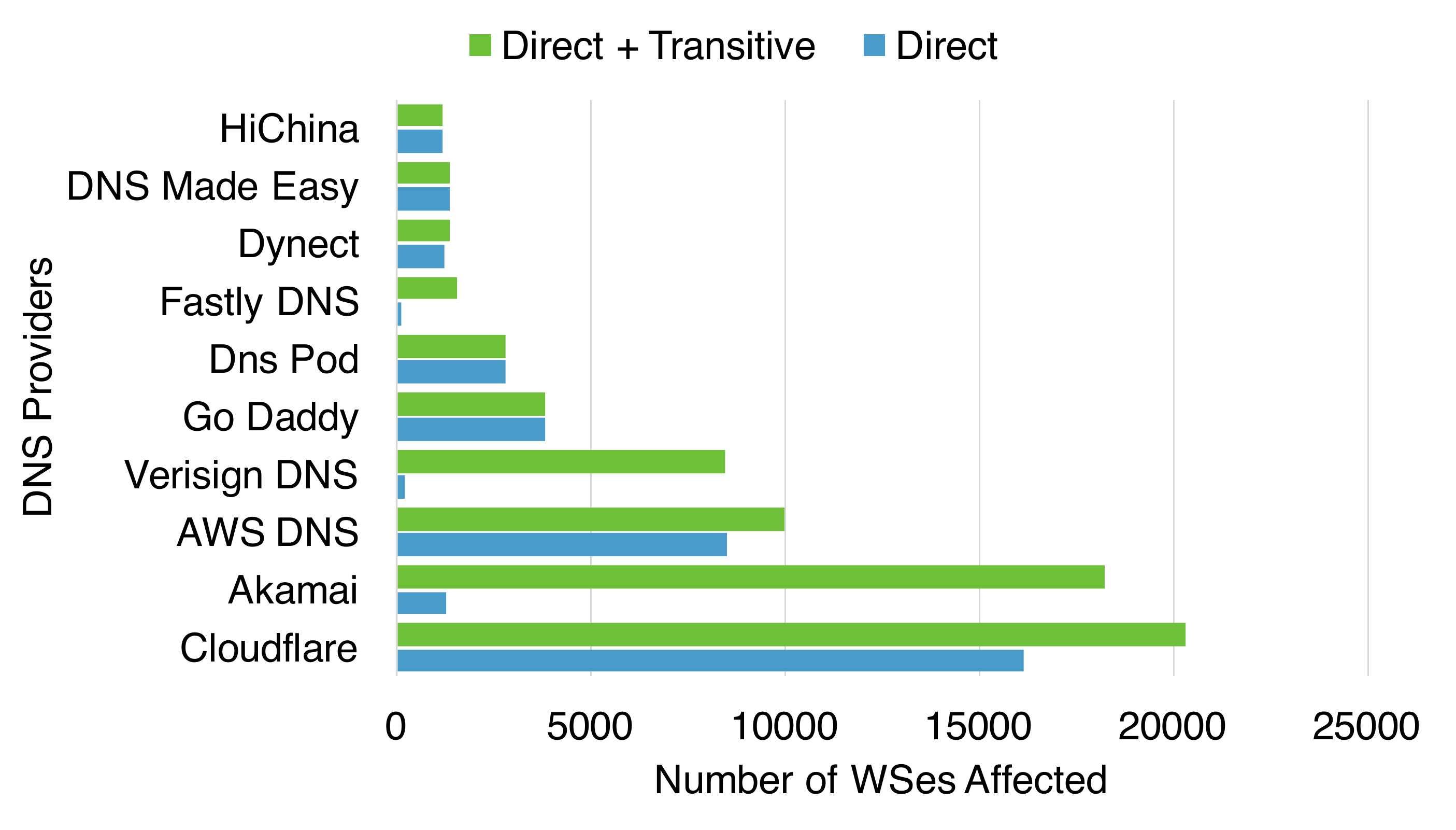}
\caption{The figure shows top 10 \Thirdpartyservice DNSs and the number of \Webservices that get affected as a result of direct, direct+transitive dependencies on these providers.} 
\label{fig:collDNS}
\end{figure}

Figure \ref{fig:collDNS} shows the top 10 DNS providers after considering indirect dependencies. It also shows their respective impact with only direct dependencies.  The x-axis shows the total number of \WSES affected. There are two interesting observations from the figure. First, some DNS providers almost have little to no direct effect on their failure but have significant transitive impact; e.g., Verisign DNS and Fastly DNS. These providers would not even show up in the most popular DNS providers if only direct dependencies with \WSES were considered. Second, for almost all of these top DNS providers, the transitive impact on them being unavailable affects significantly more \WSES than those that directly depend on them. For example, Cloudflare would adversely affect 20.32\% \WSES when including indirect impact, as compared to 16\% \WS directly. Similarly, Akamai directly affects 1.3\% of \Webservices. However, with indirect dependency, it can affect 18\% of the total \Webservices which is more than 10X amplification.

\begin{figure}[!t]
\centering
\includegraphics[width=1\columnwidth]{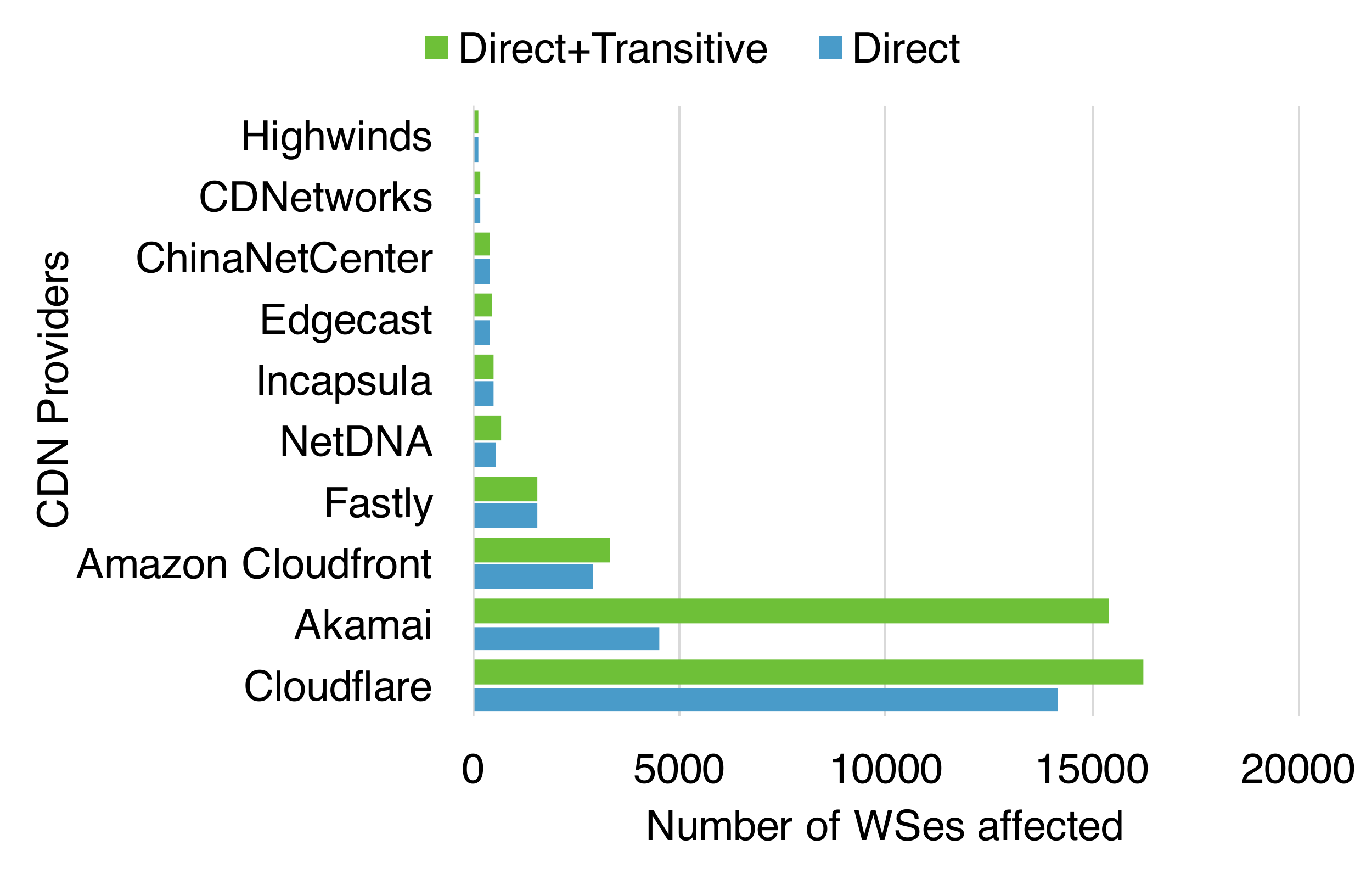}
\caption{The figure shows top 10 \Thirdpartyservice CDNs and the number of \Webservices that get affected as a result of direct, direct+transitive dependencies on these providers.} 
\label{fig:collcdn}
\end{figure}

Figure \ref{fig:collcdn} similarly compares the number of \Webservices affected for the top 10 CDN providers, with and without considering transitive dependencies. Similar to the case for third party DNS providers, if some providers (e.g. Akamai and Cloudflare) went down they would adversely affect significantly more \WSES \textit{indirectly}. Akamai being unavailable would affect 15\% of \WSES when including indirect impact, as compared to 4.5\%

\paragraph{Implications}  We conjecture that some of these providers that are not in the top-k for direct dependencies  but appear in the top-k coverage via indirect dependencies are likely less provisioned. This has key implications from an attack perspective as the attackers in the future may be able to have the same volume of impact with significantly fewer resources. This also points to a natural direction for future work---we need to explore the capacity and hosting footprint of these services in greater depth. Also, this information is useful if an attacker wants to amplify a particular kind of attack, for example, consider if the attacker's goal is to do an RoS attack while maximizing damage, then the attacker could attack an indirect link on which many OCSPs depend etc.

\begin{insight}
Targeting low-ranked \Webservices can cause a collateral damage of upto 20\% to the top 100k \Webservices
\label{insight:attackSurface}
\end{insight}

Finally, we consider a form of {\em collateral damage} that could result due to indirect dependencies. Specifically, we consider the collateral damage can be defined as the number of \Webservices affected as a result of targeting one particular \Webservice. 
  That is, imagine a \Webservice W being subject to a \DoS attack.
   Now, suppose this attack   impacts the performance and availability of the \Thirdpartyservice T that W uses. The impact on T in turn 
   can cause other \Webservices W' to also be indirectly impacted 
    because of this attack. We acknowledge that this type of analysis is speculative and there are many caveats here as we assume that T does not have perfect resource isolation across its various customers. We argue that this is in fact quite plausible as not all resources (E.g., network bandwidth inside a datacenter or WAN connectivity from an exchange point) may be multiplexed and there may less-than-perfect isolation. Thus, from a defensive perspective this is a potential attack we need to consider in the future.

For this analysis, we consider the  six different attack vectors  to attack a \Webservice: three of these arise as a result of direct dependency of \Webservices on CDNs, DNS and OCSP servers. The other three arise as a result of dependencies among \Thirdpartyservices, and hence, a \Webservice can attacked from an indirect link as well.
 
\begin{figure}[!t]
\centering
\includegraphics[width=1\columnwidth]{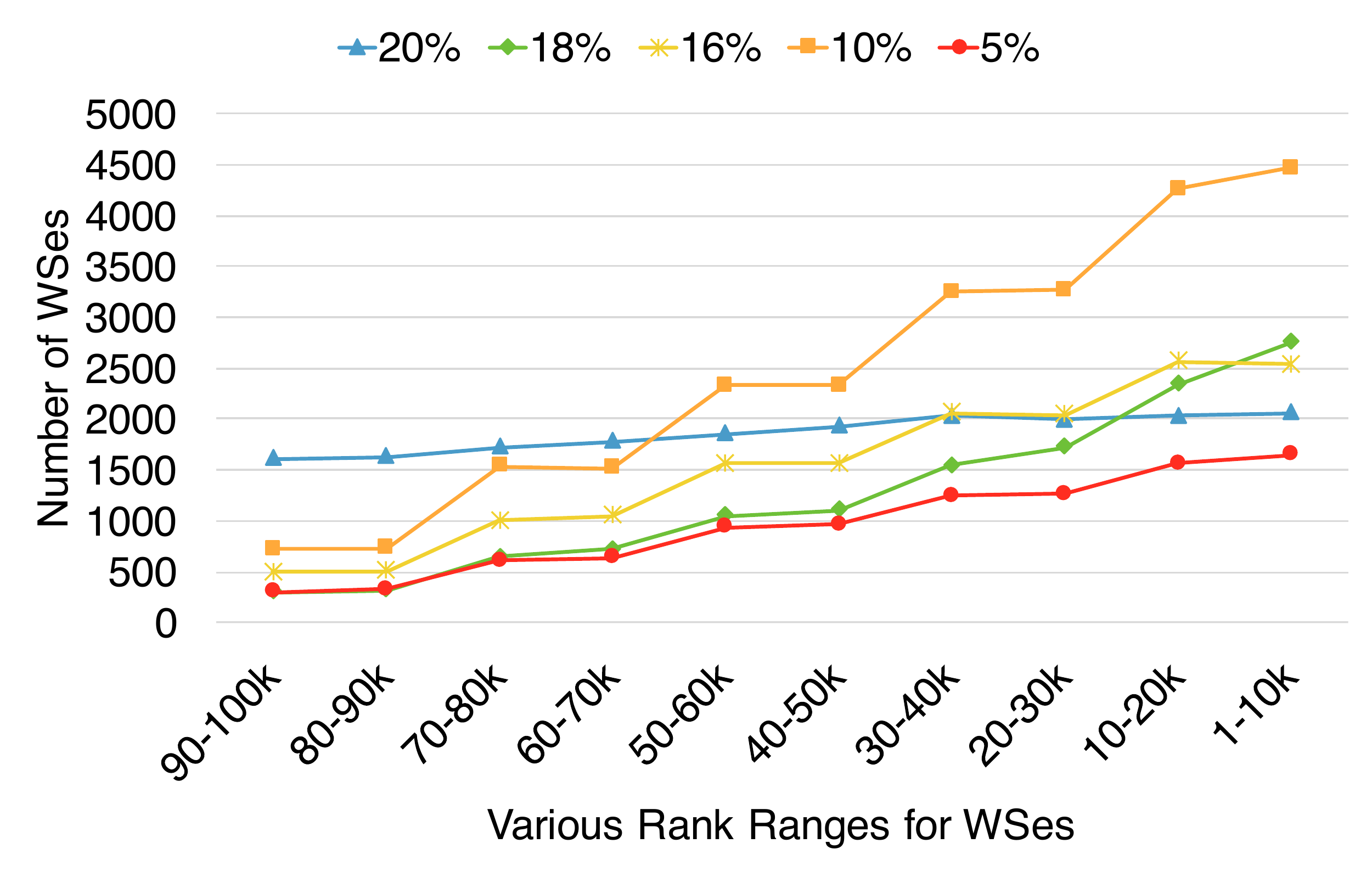}
\caption{ Distribution of \Webservices across ranks for the five maximum \%age collateral damage observed.  In the rank range from 90-100k \Webservices, there exist about 1500 \Webservices that can cause a \%age collateral damage of 20\%} 
\label{fig:collWeb}
\end{figure}

Note that since there are a few major \Thirdpartyservices, the collateral damage of all \Webservices dependent on a particular provider will be same.
We calculate the collateral damage associated with all the top 100k \Webservices, across all the six attack vectors and find the maximum collateral damage associated with each \Webservice. Then we rank these \Webservices based on their collateral damage.  We observed that maximum collateral damage associated with a \Webservice to be 20\% of the top 100k \Webservices. In order to understand, the rank distribution among these \Webservices, that cause a given collateral damage, we divide top 100k \Webservices into 10 groups of range 10,000 \Webservices as shown in figure \ref{fig:collWeb} and see for a given collateral damage, let's say 20\%, how many \Webservices fall in a particular range group. 
It can be seen from the graph that roughly 1500 \Webservices in the rank range of 90k-100k can cause a collateral damage of 20\% to the top 100k \WSES. The graph shows this trend for the five maximum \%age collateral damage observed.

\paragraph{Implications} 

This is important from the \Webservice point of view. Some popular \Webservices may not want to be co-dependent on a \Thirdpartyservice provider, with some low-ranked or highly susceptible to be targeted \Webservices for instance, in the past there have been attacks on gambling sites \cite{gamble}. This information can help \Webservices in making business decisions with robustness in mind.

%\subfile{webtowebdep}
\section{Discussion and Recommendations}

\label{sec:discuss}
Based on our observations and analysis, we discuss key  recommendations  for different stakeholders. 

\mypara{Web services}
The first and obvious recommendation for \Webservices is that they need to build in more resilience and redundancy when using \Thirdpartyservices. In fact, we already see this happening to some effect as a result of the Dyn incident (see Table~\ref{table:dynStats}). 
 A second, and perhaps more subtle recommendation is that \Webservices may need to analyze the hidden dependencies of the \Thirdpartyservices they use as they maybe indirectly exposed to potential threats. For instance, the types of analysis we have performed  can be made available as a neutral service that \Webservices can query before making  business decisions.

\mypara{Third party services} \Thirdpartyservices 
 (e.g., like Dyn) should be more transparent about the types of attacks they see and also about the potential redundancies they have in place to tackle large-scale attacks. Furthermore, \Thirdpartyservices should also be judicious in how they rely on other \Thirdpartyservices since the transitive dependencies can significantly impact a large fraction of \Webservices. 

\mypara{Attackers} On the negative side, our findings also have key implications for future attack strategies. The simplest observation is that the attackers can use these results and measurement techniques as a form of reconnaissance to inform their future targets to maximize their return-on-investments. For instance, the attackers may induce a larger impact by attacking less-provisioned \Thirdpartyservices rather than directly attacking well-provisioned \Webservices. Furthermore,  attackers may launch indirect attacks by targeting seemingly unrelated \Thirdpartyservices to create indirect \DoS and \RoQ attacks for targets of interest. Such indirect attacks may be hard to debug and diagnose and  for \Webservices to defend against since they may not have any direct contractual relationships with the impacted \Thirdpartyservices.

\section{Related Work}

\label{sec:related}

Our study is related to a rich literature  of measurements of the web ecosystem and the Internet infrastructure. However, there is surprisingly little or no systematic analysis of the fragility of the web ecosystem with respect to the  \Thirdpartyservices and types of direct and indirect dependencies we report  here. 

\mypara{Website complexity and performance}
 There are several prior works that have analyzed the widespread use of third-party content in popular websites. Butkiewicz et~al., study the impact of the complexity of the website as measured in terms of the number of objects and number of third-party objects on the website load time (e.g.,~\cite{butkiewicz2011understanding}). Other work  systematically analyzes the critical paths to understand if and how specific content impacts the page load time (e.g.,~\cite{wang-nsdi13}). These are related to a potential \RoQ attack; however, our focus is on the infrastructure services at a higher level than individual websites. Other work focuses on the privacy implications of the tracking services that appear on the website (e.g., ~\cite{roesner2012detecting,lerner-usenixsec16,privdiff-www09}). The focus on privacy implications is orthogonal to our work. 

%imc paper, roesner paper \\ 

\mypara{Dependency measurement in distributed systems} 
There are a number of other efforts to understand the dependencies in distributed systems in order to help with performance debugging and diagnosis (e.g.,~\cite{webprophet-nsdi10}). This type of related work is complementary to our work since they focus on the intra-hosting and internal dependencies rather than the external facing dependencies on third-party services. 

%WebProphet\\
%anything else?

\mypara{Understanding CDN and hosting}
 Given the large footprint and popular use of CDNs and hosting services, there have been a number of measurement studies to understand CDNs as early as 2001~\cite{cdn-imc01}. For instance, recent work analyzes the hosting infrastructure of popular websites~\cite{webcartography}. Other work maps the growing infrastructure and edge deployment by popular content providers (e.g., \cite{calder-imc13}).  Such analysis can complement our analysis especially as we do not consider the hosting infrastructure and capacity bottlenecks of these popular services.  This is a natural direction for future work to extend our analysis. Other recent work points out an increasing adoption of DDoS protection services by \Webservices~\cite{Jonker:2016}.

\mypara{Internet-wide measurement} 
 In addition to web measurement, there is also a broader interest in understanding Internet infrastructure. For instance, Zmap~\cite{durumeric2013zmap} and Censys~\cite{durumeric2015search} present mechanisms to scan the Internet to understand the open and vulnerable 
  services.  Our focus on web infrastructure 
   is complementary to this work. 
   
 \mypara{TLS and certificate measurements} 
 Other work has analyzed the use of TLS, the certificate ecosystem, and the use of Certificate 
    Revocation in the wild (e.g., ~\cite{chung-ssl-imc16, vandersloot-imc16,liu-ssl-imc15}). These  suggest potential \RoS attacks that could be executed via the \Thirdpartyservices we analyze here.

\section{Conclusions}

\label{sec:conclusion}
This paper was motivated by the recent Dyn incident that led to several popular web services being offline as a result of a  \DoS attack on the DNS provider. This motivated us to look at the broader landscape of the robustness of modern web services, the dependencies they have on third-party services,  as well as more hidden and transitive dependencies via these third-party services.  Our analysis paints a somewhat bleak situation on the state of modern web ecosystem:  (1) most webservices have little to no redundancy when using third-party infrastructure services (e.g., CDN, DNS, CAs); (2) a small number of thirdparty services could become the Achilles' heel for the web ecosystem; and (3) considering transitive dependencies can further exacerbate the attack surface and also suggest more opportunities to affect the quality, availability, and security of popular webservices. These observations have key implications and can lead to concrete and actionable recommendation for different stakeholders (i.e., attackers, web services, and third-party services). For instances, web and third-party services should understand their effective attack surface via direct and indirect dependencies and build sufficient levels of redundancy, while attackers can leverage these insights to optimally choose targets for maximizing their future impact.

{\footnotesize \bibliographystyle{acm}
\bibliography{sample}}

% \theendnotes

\end{document}